\documentclass[12pt]{iopart}

\usepackage{iopams,color,epsfig,xfrac,enumerate}

\begin{document}

\vspace*{-1.6cm}
\thispagestyle{empty}
\begin{flushright}
BROWN-HET-1659, WITS-CTP-146
\end{flushright}
\vspace*{0.5cm}

\title{Canonical Formulation of $O(N)$ Vector/Higher Spin Correspondence}

\author{Robert de Mello Koch$^1$, Antal Jevicki$^2$, Jo\~ao P. Rodrigues$^1$ and Junggi Yoon$^2$}

\address{$^1$National Institute for Theoretical Physics, School of Physics and Centre for Theoretical Physics, University of the Witwatersrand, Wits, 2050, South Africa}
\address{$^2$Department of Physics, Brown University, Providence, RI 02912, USA}
\ead{robert@neo.phys.wits.ac.za, antal\_jevicki@brown.edu, joao.rodrigues@wits.ac.za, jung-gi\_yoon@brown.edu}

\begin{abstract}
We discuss the canonical structure of the collective formulation of Vector Model/Higher Spin Duality in AdS$_4$. This involves a construction of bulk AdS Higher Spin fields through a time-like bi-local Map, with a Hamiltonian and canonical structure which are established to all orders in $1/N$.
\end{abstract}

\pacs{11.15.Pg, 04.62.+v}
%
%
%
%
%

\section{Introduction}
\label{sec:intro}

One of the simplest non-stringy example of AdS/CFT correspondence \cite{Maldacena:1997re,Gubser:1998bc,Witten:1998qj} is the duality between vector type models and Higher Spin Gravity in one higher dimension \cite{Klebanov:2002ja,Sezgin:2002rt}. These dualities are presently a topic of fruitful studies \cite{Giombi:2009wh,Giombi:2013fka,Jevicki:2014mfa,Koch:2014mxa}. Higher Spin theories have a long history \cite{Rarita:1941mf,Fronsdal:1978rb,Fang:1978wz}. Interacting theories of all spins containing Gravity were successfully constructed through a gauge principle \cite{Fradkin:1986ka,Vasiliev:1990en,Vasiliev:1995dn,Bekaert:2005vh}. 

A construction of Higher Spin/vector model Duality based on collective fields was proposed in \cite{Das:2003vw}. A one-to-one Map was explicitly given in \cite{Koch:2010cy,Jevicki:2011ss} in the light-cone gauge \cite{Metsaev:1999ui} (where HS gravity is the simplest). A similar identification of AdS space in light-cone QCD was developed in \cite{Brodsky:2013dca,Brodsky:2014yha}. A renormalization group method for bi-local observables is being developed in \cite{Leigh:2014tza,Leigh:2014qca, Zayas:2013qda,Douglas:2010rc}. The collective method is easily formulated in any time-like frame, and it has been employed explicitly in \cite{deMelloKoch:2012vc,Das:2012dt}. A covariant version is also possible with a more general first principle understanding of the Duality \cite{Koch:2014mxa}.
 
This concrete AdS/CFT Duality provides insight and allows constructions and studies of issues which otherwise are fairly difficult in HS Gravity and String Theory. One such issue is the procedure for quantization of the theory. In Higher Spin Gravity, Vasiliev's equations of motion are known explicitly but the canonical description of the equation has been only started in \cite{Metsaev:2011iz}. In String theory, this has not been possible at all except in very low dimension.
 
On the other hand the collective construction naturally contains a canonical picture which, through Duality, also gives the canonical picture of Higher Spin Gravity. For that reason, in this paper we discuss in detail the canonical structure brought in by the Collective/HS Gravity identification. This concerns the form of bulk Higher Spin fields and observables, their exact commutation relations and AdS locality. These will be established systematically in the remainder of this paper. Finally we stress that the present work deals with the canonical construction in a time-like gauge of the theory. This gauge is appropriate for Hamiltonian description and is central to the unitarity of the theory. It is of definite interest to have the bi-local map formulated in the covariant gauge framework also. Some elements of to map to covariant (Fronsdal type) gauges were recently given in a separate work \cite{Koch:2014mxa}. This construction uses the world line spinning particle framework and as such could be related to the bi-local higher spin holographic construction of \cite{Vasiliev:2012vf} given in terms spinor variables. This connection is currently being considered.

\section{Basics}
\label{sec:basi}

The basis of the AdS/CFT lies in the different manifestation of the theory when expanded perturbatively vs when seen through the Large $N$ expansion. Collective field theory is built to implement this picture to all orders in $1/N$ with a collective representation \cite{Jevicki:1979mb} of the Hamiltonian:
\begin{equation}
H=H_{\text{CFT}}=H_{\text{col}}\left(\Psi_c,1/N\right)
\end{equation}
which is systematically given in powers of $1/N$.
\begin{equation}
H_{\text{col}}=N H_0+H_2+\frac{1}{\sqrt{N}}H_3+\frac{1}{N}H_4+\cdots\label{Hamiltonian expansion}
\end{equation}
in terms of collective fields. The main property of collective fields is that they are canonical. i.e.
\begin{equation}
\left[\Psi_c,\Pi_{c'}\right]=\delta_{c,c'}\label{col commutation rel}
\end{equation}
Here $c$ labels the collective degrees of freedom, with kinematics that depends on the theory. For the case of $O(N)$ vector models, one has the bi-local collective fields $\Psi(t;\vec{x}_1,\vec{x}_2)$ defined as
\begin{equation}
\Psi\left(t;\vec{x}_1,\vec{x}_2\right)\equiv\sum_{i=1}^N \varphi^i\left(t,\vec{x}_1\right)\varphi^i\left(t,\vec{x}_2\right)
\end{equation}
In this case, the collective Hamiltonian is given by
\begin{equation}
\hspace*{-2cm}
H_{\text{col}}=\frac{2}{N}\Tr\left(\Pi \Psi \Pi\right)+\frac{N}{8}\Tr \Psi^{-1}+\frac{N}{2}\int d\vec{x} \left[-\nabla_x^2\left.\Psi(\vec{x},\vec{y})\right|_{\vec{y}=\vec{x}}\right]+V[\Psi]+\Delta V\label{collective hamiltonian}
\end{equation}
Here $V[\Psi]$ represents the original interaction potential and $\Delta V$ represents (known) lower order counter-terms.(A detailed expression can be seen in \cite{Jevicki:2014mfa}) The $1/N$ series is generated systematically as follows: One first determines the Large $N$ background field $\Psi_0(\vec{x}_1,\vec{x}_2)$ through minimization of the collective Hamiltonian. Expanding the bi-local field $\Psi\left(t,\vec{x}_1,\vec{x}_2\right)$ around the background field $\Psi_0(\vec{x}_1,\vec{x}_2)$
\begin{equation}
\Psi\left(t;\vec{x}_1,\vec{x}_2\right)=\Psi_0\left(\vec{x}_1,\vec{x}_2\right)+\frac{1}{\sqrt{N}}\Psi'\left(t;\vec{x}_1,\vec{x}_2\right)
\end{equation}
(\ref{collective hamiltonian}) gives the series of higher vertices interaction vertices:
\begin{equation}
H_n=\Tr(\underbrace{\Psi_0^{-1}\star\Psi'\star\cdots\star\Psi_0^{-1}\star\Psi'}_{n}\star\Psi_0^{-1})
\end{equation}
with a natural star product defined as $A\star B\equiv\int d\vec{x}_2\;A(\vec{x}_1,\vec{x}_2)B(\vec{x}_2,\vec{x}_3)$ representing a matrix product in bi-local space. Here we assumed the scalar field interaction $V$ to be at most quartic in $\varphi$, otherwise there is an additional interaction term generated from $V$. To summarize, the two main properties of the collective representation are:

\begin{enumerate}[I.]
\item The representation features $1/N$ as a coupling constant, leading to a natural Witten type expansion.

\item Collective fields are exactly canonical, through the commutation relations (\ref{col commutation rel}).
\end{enumerate}

This property follows from the definition of the conjugate field $\Pi={\partial \over \partial \Psi}$. For the bi-local case we therefore have
\begin{equation}
[\Psi(x_1,x_2),\Pi(x'_1,x'_2)]=\delta(x_1-x_1')\delta(x_2-x_2')+\delta(x_1-x_2')\delta(x_2-x_1')
\end{equation}
for all order in $1/N$. The only nontrivial issue with respect to exact duality is the interpretation of the `collective space' $c$. This represents a kinematical problem. When interpreted in physical terms, the collective space leads to extra emerging coordinates (as in \cite{Das:1990kaa}) and emerging gravitational and string degrees of freedom (\cite{deMelloKoch:2003pv}). In some relatively simple cases, this `de-coding' of collective degrees can be completely done. Such is the case of $N$-component vector theories with dual Higher Spin fields and AdS$_{d+1}$. For these one has a one-to-one map between bi-local (collective) degrees of freedom and Higher Spin fields in AdS space-time. 

The simplest form of the Map is found in light-cone quantization where it is given explicitly for AdS$_4$/CFT$_3$ in \cite{Koch:2010cy}. We give a short summary of the light-cone case since the construction in any other frame will be related. Denoting a sequence of higher spin fields by $\mathcal{H}_s(x^+;x^-,x,z)$, a Fourier transformation with respect to spin $s$ leads to
\begin{equation}
\mathcal{H}(x^+;x^-,x,z,\theta)=\sum_{s=0\text{ : even} }^\infty\cos (s\theta) \;\mathcal{H}_s(x^+;x^-,x,z)
\end{equation}
representing a field on AdS$_4\times S^1$. It was established by Metsaev that a representation \cite{Metsaev:1999ui} of $SO(2,3)$ with all spins can be built on this space-time. To construct a one-to-one map between bi-local and bulk fields, one first builds a transformation \cite{Koch:2010cy} from bi-local momentum space to the momentum space\footnote{Here $\theta$, the conjugate to spin $s$, is a coordinate for $S^1$ is treated as a momentum.} of AdS$_4\times S^1$. This is given by the following transformations:
\begin{eqnarray}
p^+&=&p_1^++p_2^+\label{light-cone transformation1}\\
p&=&p_1+p_2\label{light-cone transformation2}\\
p^z&=&p_1\sqrt{\frac{p_2^+}{p_1^+}}-p_2\sqrt{\frac{p_1^+}{p_2^+}}\label{light-cone transformation3}\\
\theta&=&2\arctan\sqrt{\frac{p_2^+}{p_1^+}}\label{light-cone transformation4}
\end{eqnarray}
This transformation induces the Map between fields through:
\begin{equation}
\fl
\widetilde{\mathcal{H}}\left(p^+,p,p^z,\theta\right)=\int dp_1^+ dp_1 dp_2^+ dp_2 \;\mathcal{K}\left(p^+,p,p^z,\theta;p_1^+,p^1,p_2^+,p_2\right)\widetilde{\Psi}\left(p_1^+,p_1,p_2^+,p_2\right)
\end{equation}
where the kernel $\mathcal{K}$ is simply
\begin{eqnarray}
\hspace*{-2cm}
&&\mathcal{K}\left(p^+,p,p^z,\theta;p_1^+,p^1,p_2^+,p_2\right)\cr
\hspace*{-2cm}&=&\delta\left(p_1-\frac{\left(1+\cos\theta\right)p+\sin\theta p^z}{2}\right)\delta\left(p_2-\frac{\left(1-\cos\theta\right)p-\sin\theta p^z}{2}\right)\cr
\hspace*{-2cm}&&\times\delta\left(p_2^+-p^+\sin^2\frac{\theta}{2}\right)\delta\left(p_1^+-p^+\cos^2\frac{\theta}{2}\right)
\end{eqnarray}
This is built through the inverses of the above (momentum space) transformations. Consequently, this map between fields is one-to-one and invertible. 

The coordinate space map can be determined from this kernel through the the chain rule. In particular,
\begin{equation}
z\;\widetilde{\mathcal{H}}=-\frac{\partial}{\partial p^z}\int dp_1^+dp_1 dp_2^+ dp_2 \;\mathcal{K}\;\widetilde{\Psi}\left(p_1^+,p_1,p_2^+,p_2\right)
\end{equation}
establishes the following identification between the extra AdS$_4$ coordinate and the relative bi-local space:
\begin{equation}
z=\frac{(x_1-x_2)\sqrt{p_1^+p_2^+}}{p_1^++p_2^+}
\end{equation}
By construction, the transformations between coordinates/momenta corresponds to a canonical transformation. This fact has a number of consequences.

One can show that the $SO(2,3)$ generators of AdS$_4$ higher spin fields \cite{Metsaev:1999ui} correspond to those of the bi-local CFT$_3$ \cite{Koch:2010cy}, namely:
\begin{equation}
L^{AdS}\widetilde{\mathcal{H}}(p^+,p,p^z,\theta)=\int dp_1^+ dp_1dp_2^+ dp_2\;\mathcal{K}\; L^{\text{bi-local}}\widetilde{\Psi}(p_1^+,p_1,p_2^+,p_2)
\end{equation}
for any generator of $SO(2,3)$. Furthermore, the canonical commutators of collective fields are seen to imply canonical commutation relations in the constructed AdS$_4$ space. Canonical quantization of scalar field in light-cone time introduces the notation $\widetilde{\varphi}^i(p^+,x)$ which is Fourier transformed with respect to $x^-$. For $p^+>0$, $\widetilde{\varphi}^i(p^+,x)$ and its conjugate $\widetilde{\varphi}^i(-p^+,x)$ plays a role of creation operator and annihilation operator, respectively. We define bi-local field to be
\begin{equation}
\widetilde{\Psi}(p^+_1,x_1,p^+_2,x_2)={1\over \sqrt{N}}\widetilde{\varphi}^i(p^+_1,x_1)\widetilde{\varphi}^i(p^+_2,x_2)
\end{equation}
And, one can represent its conjugate as
\begin{equation}
\fl
\widetilde{\Psi}^\dag(p^+_1,x_1,p^+_2,x_2)={1\over \sqrt{N}}\widetilde{\varphi}^i(-p^+_1,x_1)\widetilde{\varphi}^i(-p^+_2,x_2)={p_1^+p_2^+\over \sqrt{2N}} {\partial \over \partial \widetilde{\varphi}^i(p_1^+,x_1)}{\partial \over \partial \widetilde{\varphi}^i(p_2^+,x_2)}
\end{equation}
Using chain rule, one can easily express $\widetilde{\Psi}^\dag(p^+_1,x_1,p^+_2,x_2)$ in terms of $\widetilde{\Psi}(p^+_1,x_1,p^+_2,x_2)$ and its conjugate ${\partial \over \partial \widetilde{\Psi}(p_I^+,x_I,p_J^+,x_J)}$.
\begin{equation}
\hspace*{-2cm}
\widetilde{\Psi}^\dag(p_1^+,x_1,p_2^+,x_2)=p_1^+p_2^+{\partial\over \partial \widetilde{\Psi}(1;2)}+{p_1^+p_2^+\over N}\sum_{I,J}\widetilde{\Psi}(I,J){\partial \over \partial \widetilde{\Psi}(J,1)}{\partial \over \partial \widetilde{\Psi}(2,I)}\label{eq:psidag}
\end{equation}
up to contact terms and where we used compact notation, $I=(p_I^+,x_I)$. Note that the contact terms in (\ref{eq:psidag}) appear because one has to take independent collective fields $\widetilde{\Psi}$ into account when using chain rule. The algebraic structure and $1/N$ expansion of these $O(N)$ bi-local operators was studied in \cite{deMelloKoch:2012vc}. We also present this algebraic structure in the $U(N)$ example in \ref{sec:algebra of bi-local operators}. The derivative defines a canonical conjugate of the bi-local field $\widetilde{\Psi}(p_1^+,p_1,p_2^+,p_2)$, and we denote it by $\widetilde{\Pi}(p_2^+,p_2,p_1^+,p_1)$ with the canonical (bi-local) commutation relation:
\begin{eqnarray}
\fl
\left[\widetilde{\Phi}(p_1^+,p_1,p_2^+,p_2),\widetilde{\Pi}(k_2^+,k_2,k_1^+,k_1)\right]&=&{ip_1^+p_2^+\over 2} \delta(p_1^+-k_1^+)\delta(p_1-k_1)\delta(p_2^+-k_2^+)\delta(p_2-k_2)\cr
\fl
&&+( (k_1^+,k_1)\;\longleftrightarrow\; (k_2^+,k_2))
\end{eqnarray}
It follows that the Higher Spin fields $\widetilde{\mathcal{H}}(p^+,p,p^z,\theta)$ and $\widetilde{\mathcal{W}}(p^+,p,p^z,\theta)$ constructed from $\widetilde{\Psi}$ and $\widetilde{\Pi}$ through the kernel $\mathcal{K}(p^+,p,p^z,\theta;p_1^+,p^1,p_2^+,p_2)$:
\begin{eqnarray}
\widetilde{\mathcal{H}}(p^+,p,p^z,\theta)&=&\int dp_1^+ dp_1dp_2^+ dp_2\;\mathcal{K}\;\widetilde{\Phi}\\
\widetilde{\mathcal{W}}(p^+,p,p^z,\theta)&=&\int dp_1^+ dp_1dp_2^+ dp_2\;\mathcal{K}\;\widetilde{\Pi}
\end{eqnarray}
obey canonical commutation relations:
\begin{equation}
\hspace*{-2cm}
\left[\widetilde{\mathcal{H}}(p^+,p,p^z,\theta),\widetilde{\mathcal{W}}(k^+,k,k^z,\phi)\right]=ip^+\delta(p^+-k^+)\delta(p-k)\delta(p^z-k^z)\delta(\theta-\phi)
\end{equation}

Continuing with the description of the construction, we emphasize the feature that as constructed it is off-shell. Equations of motion and time evolution of the dual theory will be dictated by those of bi-local collective fields $\widetilde{\Psi}$. They are given to all orders in $1/N$ by the Hamiltonian in (\ref{Hamiltonian expansion}). In leading expansion, the (linearized) equations for the bi-local field read:
\begin{equation}
\left({1\over i}{\partial\over\partial x^+}+\frac{p_1^2}{2p_1^+}+\frac{p_2^2}{2p_2^+}\right)\widetilde{\Psi}(x^+;p_1^+,p_1,p_2^+,p_2)=0
\end{equation}
coming from the leading term in the $1/N$ expansion of the Hamiltonian (\ref{Hamiltonian expansion}). This gives the time evolution:
\begin{equation}
\widetilde{\Psi}(x^+;p_1^+,p_1,p_2^+,p_2)=e^{ix^+\left(-\frac{p_1^2}{2p_1^+}-\frac{p_2^2}{2p_2^+}\right)}\widetilde{\Psi}(p_1^+,p_1,p_2^+,p_2)
\end{equation}
and after the momentum space map, the following leading time evolution for the Higher Spin field $\widetilde{\mathcal{H}}(x^+;p^+,p,p^z,\theta)$:
\begin{equation}
\widetilde{\mathcal{H}}(x^+;p^+,p,p^z,\theta)=e^{-ix^+\frac{p^2+(p^z)^2}{2p^+}}\widetilde{\mathcal{H}}(p^+,p,p^z,\theta)
\end{equation}
and the associated linearized equation of motion:
\begin{equation}
\left(p^+\frac{1}{i}\frac{\partial}{\partial x^+}+p^2+(p^z)^2\right)\widetilde{\mathcal{H}}(x^+;p^+,p,p^z,\theta)=0
\end{equation}
agree with the Higher Spin equations of Metsaev \cite{Metsaev:1999ui} .

Let us now come to another relevant property of this bulk construction, which is the behavior of the constructed AdS fields near the $z=0$ boundary. For this, we can proceed to be on-shell and in linearized approximation, and consider the map for the spin $s$ field in particular:
\begin{eqnarray}
\fl\mathcal{H}_s(x^+;x^-,x,z)&=&\int d\theta\; \mathcal{H}(x^+;x^-,x,z,\theta)\cos (s\theta )\cr
\fl&=&\int  d^4p\; \delta (2p^+p^-+p^2+ (p^z )^2 )e^{ix^\mu p_\mu}\int dp^+_1dp_1 dp^+_2dp_2\;\mathcal{J}(p_1^+,p_2^+)\cr
\fl&&\times\delta (p^+_1+p^+_2-p^+ )\delta (p_1+p_2-p )\delta \left(p_1\sqrt{\frac{p^+_2}{p^+_1}}-p_2\sqrt{\frac{p^+_1}{p^+_2}}-p^z\right)\cr
\fl&&\times\frac{s!}{\Gamma\left(s+\frac{1}{2}\right)} P_s^{-\frac{1}{2},-\frac{1}{2}}\left(\frac{p^+_2-p^+_1}{p^+_2+p^+_1}\right)\widetilde{\Psi}(p^+_1,p_1,p^+_2,p_2)
\end{eqnarray}
where $\mathcal{J}=\left|\frac{\partial\left(p^+,p,p^z,\theta\right)}{\partial\left(p_1^+,p_1,p_2^+,p_2\right)}\right|=\frac{1}{p_1^+}+\frac{1}{p_2^+}$ is Jacobian of the transformation in (\ref{light-cone transformation1})$\sim$(\ref{light-cone transformation4}) and $P_s^{-\frac{1}{2},-\frac{1}{2}}(x)$ is Jacobi polynomial. We also note that when put on-shell through linearized approximation the Map produces the low spin formulae of \cite{Bena:1999jv,Hamilton:2005ju,Heemskerk:2012np,Kabat:2012hp,Aizawa:2014yqa}.

Using the symmetry of the collective bi-local field $\widetilde{\Psi}(p^+_1,p_1,p^+_2,p_2)=\widetilde{\Psi}(p^+_2,p_2,p^+_1,p_1)$ which corresponds to a symmetry of the higher spin field $\widetilde{\mathcal{H}}(p^+,p,p^z,\theta)=\widetilde{\mathcal{H}}(p^+,p,-p^z,\pi-\theta)$ by the kernel $\mathcal{K}$ and performing a change of variables from $p^z$ to $p^-=-\frac{p^2+(p^z)^2}{2p^+}$ (coming from the delta function), one obtains, after short calculation, the relation:
\begin{eqnarray}
\hspace*{-2cm}
\mathcal{H}_s(x^+,x^-,x,z)&=&\int_{-2p^+p^--p^2>0}\! dp^- dp^+ dp \; e^{ix^+ p^-+ix^-p^++ixp}J_{-\frac{1}{2}}\left(z\sqrt{-2p^+p^--p^2}\right)\cr
&&\times\sqrt{\frac{\pi z\sqrt{-2p^+p^--p^2} }{2}}\frac{s!}{\Gamma\left(s+\frac{1}{2}\right)\left(p^+\right)^{s}}\widetilde{\mathcal{O}}_s(p^-,p^+,p)
\end{eqnarray}
where $\widetilde{\mathcal{O}}_s(p^-,p^+,p)$ is recognized to represent Fourier modes of spin-$s$ primary operators $\mathcal{O}_{--\cdots -}(x)$ of the $O(N)$ model \cite{Makeenko:1980bh,Mikhailov:2002bp} which in terms of the bi-local collective read:
\begin{eqnarray}
\hspace*{-2cm}
\widetilde{\mathcal{O}}_s(p^-,p^+,p)&=&\int dp^+_1dp_1 dp^+_2dp_2\;\mathcal{J}(p_1^+,p_2^+)\cr
&&\times\delta(p^+_1+p^+_2-p^+)\delta\left(-\frac{p_1^2}{2p^+_1}-\frac{p_2^2}{2p^+_2}-p^-\right)\delta\left(p_1+p_2-p\right)\cr
&&\times\left(p^+_1+p^+_2\right)^s P_s^{-\frac{1}{2},-\frac{1}{2}}\left(\frac{p^+_2-p^+_1}{p^+_2+p^+_1}\right)\widetilde{\Psi}(p^+_1,p^+_2,p_1,p_2)\qquad
\end{eqnarray}
Now one registers the following boundary behavior of $\mathcal{H}_s(x^+,x^-,x,z)$:
\begin{equation}
\hspace*{-2cm}
\mathcal{H}_s(x^+,x^-,x,z) \stackrel{z\rightarrow 0}{\longrightarrow} \frac{s!}{\Gamma\left(s+\frac{1}{2}\right)\partial_-^s}\int_{-2p^+p^--p^2>0} \hspace*{-1.5cm}dp^-dp^+dp\;e^{ix^+ p^-+ix^-p^++ixp} \widetilde{\mathcal{O}}_s(x^-,x^+,x)
\end{equation}
The fact that the collective field map to AdS fields leads, at $z=0$, to the conserved primary operators of the CFT is a clear verification of the above bulk construction. This represents a significant consistency check.

\section{Time-like Frame}
\label{sec:time-like frame}

We can follow the basics of the light-cone construction and give an analogous construction in any other frame. Here we give the details of quantization performed in the time-like frame. Some parts of this construction have appeared before\footnote{In particular, \cite{deMelloKoch:2012vc} where higher order calculations were done in the time-like collective method.}, while some parallel the light-cone case. Nevertheless, the full construction is sufficiently nontrivial so that we find it worthwhile to present it.

Regarding the change of (phase space) coordinates from bi-local to AdS, first it is natural to identify the center of momentum of the bi-local space with the momentum of AdS$_4$ space.
\begin{eqnarray}
\vec{p}&=&\vec{p}_1+\vec{p}_2\label{map for p}\\
p^0&=&\left|\vec{p}_1\right|+\left|\vec{p}_2\right|
\end{eqnarray}

Second, based on light-cone case, we can make use of the same ``on-shell'' condition adopted for time-like kinematics. This gives the following identification of $p^z$ with bi-local momenta:
\begin{equation}
p^z=\pm\sqrt{(p^0)^2-\vec{p}^2}=2\sqrt{\left|\vec{p}_1\right|\left|\vec{p}_2\right|}\sin\left({\varphi_1-\varphi_2\over 2}\right)\label{time-like on-shell condition}
\end{equation}
where
\begin{equation}
\vec{p}_1=\left(\left|\vec{p}_1\right|\cos\varphi_1,\left|\vec{p}_1\right|\sin\varphi_1\right)\;,\quad\vec{p}_2=\left(\left|\vec{p}_2\right|\cos\varphi_2,\left|\vec{p}_2\right|\sin\varphi_2\right)
\end{equation}
There is an ambiguity in choosing the sign of $p^z$. We determined $p^z$ in a way that the sign of $p^z$ is changed when one exchanges $\vec{p}_1$ and $\vec{p}_2$. Also, note that $\left|p^z\right|=\sqrt{2\left|\vec{p}_1\right| \left|\vec{p}_2\right|-2\vec{p}_1\cdot\vec{p}_2}$. 

Next, for further identification it is useful to use the second-order Casimir which, for the unitary irreducible representations $D(E_0,s)$ of $SO(2,3)$, equals:
\begin{equation}
\mathcal{C}_{SO(2,3)}=E_0(E_0-3)+s(s+1)
\end{equation}
where $E^0$ is the lowest energy and $s$ is the spin. The massless representations are characterized by $E_0=s+1$. At semiclassical level, ignoring the ordering term (e.g. $s+1\sim s$) the second-order Casimir for the massless representation is then
\begin{equation}
\mathcal{C}_{SO(2,3)}=2s^2=2 (p^\theta)^2
\end{equation}
where $p^\theta$ plays the same role as in the light-cone case, namely labeling the internal spin degree of freedom. Comparing this the second-order Casimir with the one expressed in terms of bi-local $SO(2,3)$ generators, one can express $p^\theta$ in terms of the bi-local variables.\footnote{Also, we chose the sign of each term to give correct Poisson brackets.}
\begin{equation}
\hspace*{-1cm}
p^\theta=\sqrt{\left|\vec{p}_1\right| \left|\vec{p}_2\right|}\cos{\varphi_1+\varphi_2\over2}\;(x_2^1-x_1^1)+\sqrt{\left|\vec{p}_1\right| \left|\vec{p}_2\right|}\sin{\varphi_1+\varphi_2\over2}\;(x_2^2-x_1^2)
\end{equation}

Finally, consider the following ansatz for $\theta$ conjugate to $p^\theta$.
\begin{equation}
\theta=\arctan\left(\frac{2\vec{p}_2\times\vec{p}_1}{(\left|\vec{p}_2\right|-\left|\vec{p}_1\right|) p^z}\right)\label{ansatz for theta}
\end{equation}
where $\vec{p}_2\times\vec{p}_1\equiv p_2^1 p_1^2-p_2^2p_1^1$. One can confirm that $\theta$ and $p^\theta$ satisfy canonical Poisson bracket
\begin{equation}
\left\{p^\theta,\theta\right\}=1
\end{equation}
and the Poisson brackets with others vanish. One can also obtain this identification of $\theta$ from a polarization vector and a primary operator. (See \ref{sec:reduction})

In sum, the identification between the bi-local (momentum) space and AdS$_4\times S^1$ (momentum) space is
\begin{eqnarray}
\vec{p}&=&\vec{p}_1+\vec{p}_2\label{time-like momentum map1}\\
p^z&=&2\sqrt{\left|\vec{p}_1\right| \left|\vec{p}_2\right|}\sin\left({\varphi_1-\varphi_2\over 2}\right)\label{time-like momentum map2}\\
\theta&=&\arctan\left(\frac{2\vec{p}_2\times\vec{p}_1}{(\left|\vec{p}_2\right|-\left|\vec{p}_1\right|) p^z}\right)\label{time-like momentum map3}
\end{eqnarray}
One can consider this identification as a point transformation in momentum space.\footnote{Again, though $\theta$ is a coordinate for $S^1$, we treat it like momentum.} The Jacobian of the transformation is
\begin{equation}
\mathcal{J}(\vec{p}_1,\vec{p}_2)=\frac{1}{\left|\vec{p}_1\right|}+\frac{1}{\left|\vec{p}_2\right|}
\end{equation}
Moreover, inverting this transformation one can express bi-local momentum in terms of momenta of AdS$_4\times S^1$ space.
\begin{equation}
\vec{p}_a=\vec{\beta}_a(\vec{p},p^z,\theta)\qquad(a=1,2)
\end{equation}
where $\vec{\beta}_a(\vec{p},p^z,\theta)$ is given in \ref{sec:inverse transformation}.
Using the momentum transformation, one can construct a higher spin field $\widetilde{\mathcal{H}}(\vec{p},p^z,\theta)$ from a bi-local field $\widetilde{\Psi}(\vec{p}_1,\vec{p}_2)$.
\begin{eqnarray}
\widetilde{\mathcal{H}}(\vec{p},p^z,\theta)&=&\int d\vec{p}_1 d\vec{p}_2\; \mathcal{K}(\vec{p},p^z,\theta;\vec{p}_1,\vec{p}_2)\widetilde{\Psi}(\vec{p}_1,\vec{p}_2)\quad\label{time-like map1}
\end{eqnarray}
where a kernel $\mathcal{K}(\vec{p},p^z,\theta;\vec{p}_1,\vec{p}_2)$ is defined to be
\begin{eqnarray}
\fl
\mathcal{K}(\vec{p},p^z,\theta;\vec{p}_1,\vec{p}_2)&=&\mathcal{J}(\vec{p}_1,\vec{p}_2)\delta^{(2)}(\vec{p}_1+\vec{p}_2-\vec{p})\delta\left(2\sqrt{\left|\vec{p}_1\right| \left|\vec{p}_2\right|}\sin\left({\varphi_1-\varphi_2\over 2}\right)-p^z\right)\cr
&&\times\delta\left(\arctan\left(\frac{2\vec{p}_2\times\vec{p}_1}{(\left|\vec{p}_2\right|-\left|\vec{p}_1\right|) p^z}\right)-\theta\right)\cr
&=&\delta^{(2)}(\vec{p}_1-\vec{\beta}_1(\vec{p},p^z,\theta))\delta^{(2)}(\vec{p}_2-\vec{\beta}_2(\vec{p},p^z,\theta))\label{time-like kernel1}
\end{eqnarray}
One can also invert the map (\ref{time-like map1}).
\begin{equation}
\widetilde{\Psi}(\vec{p}_1,\vec{p}_2)=\int  d\vec{p} dp^z d\theta\; \mathcal{Q}(\vec{p}_1,\vec{p}_2;\vec{p},p^z,\theta)\widetilde{\mathcal{H}}(\vec{p},p^z,\theta)\qquad
\end{equation}
where the inverse kernel $\mathcal{Q}(\vec{p}_1,\vec{p}_2;\vec{p},p^z,\theta)$ is
\begin{eqnarray}
\hspace*{-1cm}\mathcal{Q}(\vec{p}_1,\vec{p}_2;\vec{p},p^z,\theta)&=&\delta^{(2)}(\vec{p}_1+\vec{p}_2-\vec{p})\delta\left(2\sqrt{\left|\vec{p}_1\right| \left|\vec{p}_2\right|}\sin\left({\varphi_1-\varphi_2\over 2}\right)-p^z\right)\cr
&&\times\delta\left(\arctan\left(\frac{2\vec{p}_2\times\vec{p}_1}{(\left|\vec{p}_2\right|-\left|\vec{p}_1\right|) p^z}\right)-\theta\right)
\end{eqnarray}
The most important property of this map is again that it preserves the canonical commutation relations but now with AdS kinematics. 

Consider the bi-local field $\widetilde{\Psi}$ and its conjugate $\widetilde{\Pi}=-i{\partial \over \partial \widetilde{\Psi}}$ in the momentum space satisfying the canonical commutation relation.
\begin{eqnarray}
\left[\widetilde{\Psi}(\vec{p}_1,\vec{p}_2),\widetilde{\Pi}(\vec{k}_1,\vec{k}_2)\right]=i\left|\vec{p}_1\right|\left|\vec{p}_2\right|\delta^{(2)}(\vec{p}_1-\vec{k}_1)\delta^{(2)}(\vec{p}_2-\vec{k}_2)
\end{eqnarray}
where $(\vec{p}_1,\vec{p}_2)$ and $(\vec{k}_1,\vec{k}_2)$ are independent collective degrees of freedom. e.g. $p_1^1>p_2^1$ or $p_1^2\geqq p_2^2$ if $p_1^1=p_2^1$, and similar for $(\vec{k}_1,\vec{k}_2)$. Through the kernel $\mathcal{K}(\vec{p},p^z,\theta;\vec{p}_1,\vec{p}_2)$, one can construct $\widetilde{\mathcal{H}}(\vec{p},p^z,\theta)$ and $\widetilde{\mathcal{W}}(\vec{p},p^z,\theta)$ from $\widetilde{\Psi}$ and $\widetilde{\Pi}$, respectively.
\begin{eqnarray}
\widetilde{\mathcal{H}}(\vec{p},p^z,\theta)\!&=&\!\int\! d\vec{p}_1 d\vec{p}_2 \;\mathcal{K}(\vec{p},p^z,\theta;\vec{p}_1,\vec{p}_2)\widetilde{\Psi}(\vec{p}_1,\vec{p}_2)\\
\widetilde{\mathcal{W}}(\vec{p},p^z,\theta)\!&=&\!\int\! d\vec{p}_1 d\vec{p}_2 \;\mathcal{K}(\vec{p},p^z,\theta;\vec{p}_1,\vec{p}_2)\widetilde{\Pi}(\vec{p}_1 ,\vec{p}_2 )\quad
\end{eqnarray}
Then, $\widetilde{\mathcal{H}}(\vec{p},p^z,\theta)$ and $\widetilde{\mathcal{W}}(\vec{p},p^z,\theta)$ satisfy canonical commutation relation. i.e.
\begin{eqnarray}
\hspace*{-1cm}
\left[\widetilde{\mathcal{H}}(\vec{p},p^z,\theta),\widetilde{\mathcal{W}}(\vec{k},k^z,\phi)\right]=i\sqrt{\vec{p}^2+(p^z)^2}\;\delta^{(2)}(\vec{p}-\vec{k})\delta(p^z-k^z)\delta(\theta-\phi)
\end{eqnarray}
For $s=0$, this is the same commutation relation of scalar field in AdS$_4$ in \cite{Bena:1999jv,Fronsdal:1974ew}

Using the kernel, one can construct the transformation for the coordinates. The kernel $\mathcal{K}(\vec{p},p^z,\theta;\vec{p}_1,\vec{p}_2)$ induces the identification of bi-local coordinates $(\vec{x}_1,\vec{x}_2)$ with AdS$_4\times S^1$ space according to the chain rule. For example,
\begin{eqnarray}
\hspace*{-1cm}
x^1\widetilde{\mathcal{H}}(\vec{p},p^z,\theta)&=&-\frac{\partial}{\partial p^1}\widetilde{\mathcal{H}}(\vec{p},p^z,\theta)\cr
&=&\int d\vec{p}_1 d\vec{p}_2 \;\mathcal{K}(\vec{p},p^z,\theta;\vec{p}_1,\vec{p}_2)\left(\frac{\partial \vec{p}_1}{\partial p^1}\cdot\vec{x}_1+\frac{\partial \vec{p}_2}{\partial p^1}\cdot\vec{x}_2\right)\widetilde{\Psi}(\vec{p}_1,\vec{p}_2)
\end{eqnarray}
As a result, the map for AdS$_4\times S^1$ coordinates in terms of the bi-local variables is given by
\begin{eqnarray}
x^1&=&\frac{\left|\vec{p}_1\right|x_1^1+\left|\vec{p}_2\right|x_2^1}{\sqrt{(p^z)^2+\vec{p}^2}}-\frac{p^2 p^zp^\theta}{\vec{p}^2\sqrt{(p^z)^2+\vec{p}^2}}\label{coordiantes map1}\\
x^2&=&\frac{\left|\vec{p}_1\right|x_1^2+\left|\vec{p}_2\right|x_2^2}{\sqrt{(p^z)^2+\vec{p}^2}}+\frac{p^1 p^zp^\theta}{\vec{p}^2\sqrt{(p^z)^2+\vec{p}^2}}\label{coordiantes map2}\\
z&=&\frac{\left(\vec{x}_1-\vec{x}_2\right)\cdot \vec{p}_1\left|\vec{p}_2\right|-\left(\vec{x}_1-\vec{x}_2\right)\cdot \vec{p}_2\left|\vec{p}_1\right|}{p^z\left(\left|\vec{p}_1\right|+\left|\vec{p}_2\right|\right)}\label{coordiantes map3}\\
p^\theta&=&\sqrt{\left|\vec{p}_1\right| \left|\vec{p}_2\right|}\cos{\varphi_1+\varphi_2\over2}\;(x_2^1-x_1^1)\cr
&&+\sqrt{\left|\vec{p}_1\right| \left|\vec{p}_2\right|}\sin{\varphi_1+\varphi_2\over2}\;(x_2^2-x_1^2)\label{coordiantes map4}
\end{eqnarray}
By construction, this transformation, (\ref{time-like momentum map1})$\sim$(\ref{time-like momentum map3}) and (\ref{coordiantes map1})$\sim$(\ref{coordiantes map4}), is canonical. i.e.
\begin{equation}
\left\{p^i,x^j\right\}=\delta^{ij}\;\;\left(i,j=1,2\right)\quad\left\{p^z,z\right\}=\left\{p^\theta,\theta\right\}=1
\end{equation}
and others vanish.

The kernel $ \mathcal{K}(\vec{p},p^z,\theta;\vec{p}_1,\vec{p}_2) $ also maps the bi-local $SO(2,3)$ generators $L_{\text{bi-local}}$ to $SO(2,3)$ generators $L_{\text{ads}}$ acting on $\widetilde{\mathcal{H}}(\vec{p},p^z,\theta)$.
\begin{equation}
\int d\vec{p}_1 d\vec{p}_2\; \mathcal{K}(\vec{p},p^z,\theta;\vec{p}_1,\vec{p}_2) L_{\text{bi-local}}\widetilde{\Psi}(\vec{p}_1,\vec{p}_2)=L_{\text{ads}}\widetilde{\mathcal{H}}(\vec{p},p^z,\theta)
\end{equation}
Classically, (\ref{time-like momentum map1})$\sim$(\ref{time-like momentum map3}) and (\ref{coordiantes map1})$\sim$(\ref{coordiantes map4}) corresponds to a canonical transformation from the bi-local space to AdS$_4\times S^1$. One can also obtain $L_{\text{ads}}$ from $L_{\text{bi-local}}$ by this canonical transformation.
\begin{eqnarray}
P^\mu_{\text{ads}}&=&P^\mu_{\text{bi-local}}\label{gen identification1}\\
J^{\mu\nu}_{\text{ads}}&=&J^{\mu\nu}_{\text{bi-local}}\label{gen identification2}\\
D_{\text{ads}}&=&D_{\text{bi-local}}\label{gen identification3}\\
K^\mu_{\text{ads}}&=&K^\mu_{\text{bi-local}}\label{gen identification4}
\end{eqnarray}
Then, setting $t=0$ for simplicity, $L_{\text{ads}}$ is given by
\begin{eqnarray}
P^0_{\text{ads}}&=&\sqrt{\vec{p}^2+(p^z)^2}\label{time-likeads generator1}\\
P^1_{\text{ads}}&=&p^1\\
P^1_{\text{ads}}&=&p^2\\
J^{01}_{\text{ads}}&=& -x^1 P^0- \frac{p^2p^zp^\theta}{\vec{p}^2}\\
J^{12}_{\text{ads}}&=& x^1p^2-x^2 p^1\\
J^{20}_{\text{ads}}&=& x^2 P^0-\frac{p^1p^zp^\theta}{\vec{p}^2}\\
D_{\text{ads}}&=&x^1p^1+x^2p^2+z p^z\\
K^0_{\text{ads}}&=&-\frac{1}{2}(\vec{x}^2+z^2)P^0-\frac{p^z p^\theta }{\vec{p}^2}J^{12}-\frac{P^0(p^\theta)^2}{2\vec{p}^2}\qquad\\
K^1_{\text{ads}}&=&-\frac{1}{2}(\vec{x}^2+z^2)p^1+x^1D+\frac{zp^2P^0 p^\theta}{\vec{p}^2}+\frac{p^1(p^\theta)^2}{2\vec{p}^2}\qquad\\
K^2_{\text{ads}}&=&-\frac{1}{2}(\vec{x}^2+z^2)p^2+x^2D-\frac{z p^1P^0p^\theta}{\vec{p}^2}+\frac{p^2(p^\theta)^2}{2\vec{p}^2}\label{time-likeads generator10}\qquad
\end{eqnarray}

There is another way to obtain the map between bi-local momentum space and AdS$_4\times S^1$ momentum space. In \cite{Koch:2010cy}, the map was found by comparing generators of AdS$_4$ and CFT$_3$ in light-cone gauge. This is the inverse procedure of the derivation in this section. Unfortunately, we do not have a representation of $SO\left(2,3\right)$ for higher spin field in time-like gauge. However, there is alternative way to obtain the generators. In \cite{Metsaev:1999ui}, Metsaev constructed a realization of $SO\left(2,3\right)$ for spin-$s$ current in CFT$_3$. One can show that this agrees with a realization of $SO\left(2,3\right)$ for higher spin field in light-cone gauge after manipulating the generators. For the case of time-like gauge, one can repeat the same procedure to get $SO(2,3)$ generators in time-like gauge. (see \ref{SO(2,3) realization}, (\ref{mod gen1})$\sim$(\ref{mod gen10}) for $t=0$.) And, we can accept this result as a representation of $SO\left(2,3\right)$ for higher spin field in time-like gauge. Note that these generators in (\ref{mod gen1})$\sim$(\ref{mod gen10}) are identical to $L_{\text{ads}}$ in (\ref{time-likeads generator1})$\sim$(\ref{time-likeads generator10}). Then, one can identify $L_{\text{ads}}$ with $L_{\text{bi-local}}$ to obtain the map. i.e. see (\ref{gen identification1})$\sim$(\ref{gen identification4}).

First of all, (\ref{gen identification1}) gives expression for $\vec{p}$ and $p^z$ in terms of $\vec{p}_1$ and $\vec{p}_2$ in (\ref{time-like momentum map1}) and (\ref{time-like momentum map2}). Moreover, comparing Casimir of $SO\left(2,3\right)$ of both representation, one can get $p^\theta$ in (\ref{coordiantes map4}). Solving two equations, $J^{01}_{\text{ads}}=J^{01}_{\text{bi-local}}$ and $J^{20}_{\text{ads}}=J^{20}_{\text{bi-local}}$, for $x^1$ and $x^2$, respectively, one has (\ref{coordiantes map1}) and (\ref{coordiantes map2}). Finally, the identification $D_{\text{ads}}=D_{\text{bi-local}}$, (\ref{coordiantes map1}) and (\ref{coordiantes map2}) give the map for $z$ in (\ref{coordiantes map3}). Again, it is difficult to derive the map for $\theta$ by this identification because $\theta$ does not appear in the generators. Instead, using a primary operator and polarization vector, one can find the map for $\theta$. (see \ref{sec:reduction}). This result perfectly agrees with the previous result, (\ref{time-like momentum map1})$\sim$(\ref{time-like momentum map3}) and (\ref{coordiantes map1})$\sim$(\ref{coordiantes map4}).

\subsection{Linearization}
\label{sec:linearization}

To calculate time evolution of the bi-local collective field, one need to linearize the equation of motion by $1/N$ expansion. This procedure is standard. Minimizing the collective Hamiltonian, one determines the time-independent background field $\Psi_0(\vec{x},\vec{y})$ given by
\begin{equation}
\Psi_0(\vec{x},\vec{y})=\int d\vec{p}\; \frac{e^{i\vec{p}\cdot (\vec{x}-\vec{y})}}{2\left|\vec{p}\right|}
\end{equation}
and one expands the bi-local collective field around it: 
\begin{equation}
\Psi(t;\vec{x},\vec{y})=\Psi_0(\vec{x},\vec{y})+\frac{1}{\sqrt{N}}\Psi'(t;\vec{x},\vec{y})
\end{equation}
Note that this background field $\Psi_0\left(\vec{x},\vec{y}\right)$ equals the two-point function of the bi-local composite operator, i.e.
\begin{equation}
\Psi_0(\vec{x},\vec{y})=\left<\sum_{a=1}^N \phi^a(t;\vec{x})\phi^a(t;\vec{y})\right>
\end{equation}
Rescaling momenta $\Pi$ to get $\Pi'$ conjugate to the fluctuation $\Psi'$
\begin{equation}
\Pi(t;\vec{x},\vec{y})=\sqrt{N}\Pi'(t;\vec{x},\vec{y})
\end{equation}
one can systemically expand the Hamiltonian in large $N$.
\begin{equation}
H=N H_0+H_2+\cdots
\end{equation}
Especially, the quadratic Hamiltonian is given by
\begin{equation}
H_2=2\Tr\left(\Pi' \Psi_0 \Pi'\right)+\frac{1}{8}\Tr\left(\Psi_0^{-1} \Psi'\Psi_0^{-1} \Psi'\Psi_0^{-1} \right)+\Delta V_2
\end{equation}
where the counter term $\Delta V_2$ of order $\mathcal{O}(N^0)$ will be cancelled with the zero point energy. After Fourier transforming $\Psi'$ and $\Pi'$ and shifting them by background field $\Psi_{0,k}$ in momentum space, one has 
\begin{eqnarray}
H_2&=&\frac{1}{2}\int_{\vec{p}_1\geqq \vec{p}_2} d\vec{p}_1 d\vec{p}_2\;\left[\widetilde{\Pi}'(\vec{p}_1,-\vec{p}_2)\widetilde{\Pi}'(\vec{p}_2,-\vec{p}_1)\right.\cr
&&\qquad\qquad\qquad\qquad\left.+\omega_{\vec{p}_1,\vec{p}_2}^2\widetilde{\Psi}'(\vec{p}_1,-\vec{p}_2)\widetilde{\Psi}'(\vec{p}_2,-\vec{p}_1)\right] +\Delta V_2
\end{eqnarray}
where the frequency of the bi-local fluctuation is given by
\begin{equation}
\omega_{\vec{p}_1,\vec{p}_2}=\left|\vec{p}_1\right|+\left|\vec{p}_2\right|
\end{equation}
and the integration is performed over the independent collective degrees of freedom. Hence, the on-shell condition of the bi-local fluctuation $\widetilde{\Psi}'\left(\vec{p}_1,\vec{p}_2\right)$ is given by
\begin{equation}
\left(-\frac{\partial^2}{\partial t^2}-\left(\left|\vec{p}_1\right|+\left|\vec{p}_2\right|\right)^2\right)\widetilde{\Psi}'\left(t;\vec{p}_1,\vec{p}_2\right)=0\label{time-like on-shell condition2}
\end{equation}
and, the time evolution of the bi-local collective field $\widetilde{\Psi}'$ is
\begin{equation}
\widetilde{\Psi}' (t;\vec{p}_1,\vec{p}_2)=e^{-i(\left|\vec{p}_1\right|+\left|\vec{p}_2\right| )t}a\left(\vec{p}_1,\vec{p}_2\right)+e^{i (\left|\vec{p}_1\right|+\left|\vec{p}_2\right| )t}a^\dag(-\vec{p}_1,-\vec{p}_2)\qquad\label{time-like time evolution}
\end{equation}
where $a(\vec{p}_1,\vec{p}_2)$ and $a^\dag(\vec{p}_1,\vec{p}_2)$ are annihilation and creation operators of the bi-local field, respectively.

Like the light-cone case, the time evolution of the higher spin field follows from that of the collective field. For this, it is convenient to consider a map of annihilation and creation operators separately. i.e.
\begin{eqnarray}
\!\!\mathcal{A}(\vec{p},p^z,\theta)&=&\!\!\int \!d\vec{p}_1 d\vec{p}_2 \mathcal{K}(\vec{p},p^z,\theta;\vec{p}_1,\vec{p}_2) a(\vec{p}_1,\vec{p}_2)\\
\!\!\mathcal{A}^\dag(\vec{p},p^z,\theta)&=&\!\!\int\! d\vec{p}_1 d\vec{p}_2 \mathcal{K}(\vec{p},p^z,\theta;\vec{p}_1,\vec{p}_2) a^\dag(\vec{p}_1,\vec{p}_2)
\end{eqnarray}
Note that a transformation from $(\vec{p}_1,\vec{p}_2)$ to $(-\vec{p}_1,-\vec{p}_2)$ corresponds to a change of $(\vec{p},p^z,\theta)$ into $(-\vec{p},p^z,\theta)$ by the kernel $\mathcal{K}$. Hence, $\widetilde{\mathcal{H}}'(\vec{p},p^z,\theta)$ can be expressed in terms of $\mathcal{A}(\vec{p},p^z,\theta)$ and $\mathcal{A}^\dag(\vec{p},p^z,\theta)$ as
\begin{equation}
\widetilde{\mathcal{H}}'(\vec{p},p^z,\theta)=\mathcal{A}(\vec{p},p^z,\theta)+\mathcal{A}^\dag(-\vec{p},p^z,\theta )
\end{equation}
By construction, the quadratic Hamiltonian of the higher spin field reads
\begin{equation}
H_2\!=\!\int d\vec{p} dp^z d\theta \sqrt{\vec{p}^2+(p^z)^2}\mathcal{A}^\dag(\vec{p},p^z,\theta)\mathcal{A}(\vec{p},p^z,\theta)
\end{equation}

\subsection{Projection to Currents}\label{sec:projection to currents}

Consider Fourier transformation of $\widetilde{\mathcal{H}}'(\vec{p},p^z,\theta)$.
\begin{eqnarray}
\fl
\mathcal{H}_{\nu} (\vec{x},z)&=&\int \frac{d^2\vec{p}dp^z}{ (2\pi)^3 2\sqrt{\vec{p}^2+\left(p^z\right)^2}}\; e^{i\vec{p}\cdot \vec{x}+ip^z z}\int d\theta\;e^{- i\nu\theta} \left[\mathcal{A}(\vec{p},p^z,\theta)+\mathcal{A}^\dag(-\vec{p},p^z,\theta)\right]
\end{eqnarray}
where $\nu=\pm s$ $(s=0,1,2,\cdots)$ corresponds to two polarizations. Note that (\ref{time-like momentum map3}) implies that $\theta$ is periodic with period $\pi$. i.e. $\theta\sim \theta+\pi$. Therefore, one can deduce that $\nu$ is an even integer. Again, consider Fourier transformation of $\mathcal{A}(\vec{p},p^z,\theta)$ and $\mathcal{A}^\dag(\vec{p},p^z,\theta)$ separately. For $\nu=s$ where $s$ is a non-negative even integer,
\begin{eqnarray}
\fl
\widetilde{\mathcal{A}}_{ s}(\vec{x},z)&=&\int \frac{d^2\vec{p}dp^z d\theta\;e^{i\vec{p}\cdot \vec{x}+ip^z z-is\theta}}{(2\pi)^3 2\sqrt{\vec{p}^2+(p^z)^2}}\mathcal{A}(\vec{p},p^z,\theta)\cr
\fl&=&\int_{p^z>0} \frac{d^2\vec{p}dp^z\;e^{i\vec{p}\cdot\vec{x}}}{(2\pi)^3 2\sqrt{\vec{p}^2+(p^z)^2}}\int_{\sin {\varphi_1-\varphi_2\over 2}>0} d\vec{p}_1d\vec{p}_2\mathcal{J}(\vec{p}_1,\vec{p}_2)\cr
\fl&&\times2^{\frac{3}{2}s} (p^z)^{-s}\left[\left(\left(\epsilon_\mu p_2^\mu\right)^s+\left(\epsilon_\mu^* p_2^\mu\right)^s\right)\cos(p^z z)-i\left(\left(\epsilon_\mu p_2^\mu\right)^s-\left(\epsilon_\mu^* p_2^\mu\right)^s\right)\sin(p^z z)\right]\cr
\fl&&\times\delta(\sqrt{2p_1p_2-2\vec{p}_1\!\cdot\!\vec{p}_2}-p^z)\delta^{(2)}(\vec{p}_1+\vec{p}_2-\vec{p})
 a(\vec{p}_1,\vec{p}_2)\label{map for annihilation op}
\end{eqnarray}
where $\epsilon_\mu(\vec{p})$ and $\epsilon_\mu^*(\vec{p})$ are polarization vectors defined in \ref{sec:polarization} with $p^\mu=(|\vec{p}_1|+|\vec{p}_2|,\vec{p}_1+\vec{p}_2)$. For $\nu=-s$, $\widetilde{\mathcal{A}}_{-s}(\vec{x},z)$ is obtained from $\widetilde{\mathcal{A}}_{s}(\vec{x},z)$ by replacing $\sin(p^z z)$ with $-\sin(p^z z)$. Repeating a similar calculation for $\mathcal{A}^\dag$, one has
\begin{eqnarray}
\mathcal{H}_s(\vec{x},z)&=&\widetilde{\mathcal{A}}_s(\vec{x},z)+\left(\widetilde{\mathcal{A}}_{s}(\vec{x},z)\right)^\dag
\end{eqnarray}
and similar for $\mathcal{H}_{-s}(\vec{x},z)$. 

For boundary behavior, we introduce $\mathcal{H}_s^{(\pm)}(\vec{x},z)$
\begin{eqnarray}
\mathcal{H}_s^{(+)}(\vec{x},z)&\equiv&{1\over 2}\left[\mathcal{H}_s(\vec{x},z)+\mathcal{H}_{-s}(\vec{x},z)\right]\\
\mathcal{H}_s^{(-)}(\vec{x},z)&\equiv&{1\over 2}\left[\mathcal{H}_s(\vec{x},z)-\mathcal{H}_{-s}(\vec{x},z)\right]
\end{eqnarray}
Now, recalling the time evolution of the higher spin field induced from (\ref{time-like time evolution}), we has the time evolution of (\ref{map for annihilation op}). Then, we perform a change variable from $p^z$ to $p^0=\sqrt{\vec{p}^2+(p^z)^2}$ in the integration to find:
\begin{eqnarray}
\mathcal{H}_{ s}^{(+)}(t;\vec{x},z)&=&\frac{\sqrt{\pi} s!}{2^{\frac{3}{2}s+\frac{3}{2}}}\int_{p^0>|\vec{p}|} \frac{d^2\vec{p}dp^0}{(2\pi)^3 2p^0} e^{-ip^0t+i\vec{p}\cdot\vec{x}}z^{\frac{1}{2}}\left[(p^0)^2-\vec{p}^2\right]^{-\frac{s}{2}+\frac{1}{4}}\cr
&&\times J_{-\frac{1}{2}} (\sqrt{(p^0)^2-\vec{p}^2}\; z) 
 \left[\widetilde{\mathcal{O}}_s(p;\epsilon)+\widetilde{\mathcal{O}}_s(p;\epsilon^*)\right]+\mbox{h.c}
\end{eqnarray}
where the operators $\widetilde{\mathcal{O}}_s(p;\epsilon)$ and $\widetilde{\mathcal{O}}_s(p;\epsilon^*)$ are determined to be:
\begin{eqnarray}
\fl
\widetilde{\mathcal{O}}^s(p;\epsilon)&=&\int_{\sin {\varphi_1-\varphi_2\over 2}>0} d\vec{p}_1 d\vec{p}_2\; \mathcal{J}(\vec{p}_1,\vec{p}_2)\delta^{(3)}(p_1^\mu+p_2^\mu-p^\mu)\left(\epsilon_\mu\cdot p_2^\mu\right)^{s} \frac{2^{3s+1}}{s!}a(\vec{p}_1,\vec{p}_2)\label{physical primary operator1}\\
\fl\widetilde{\mathcal{O}}^s(p;\epsilon^*)&=&\int_{\sin {\varphi_1-\varphi_2\over 2}>0} d\vec{p}_1 d\vec{p}_2\; \mathcal{J}(\vec{p}_1,\vec{p}_2)\delta^{(3)}(p_1^\mu+p_2^\mu-p^\mu)\left(\epsilon_\mu^*\cdot p_2^\mu\right)^{s} \frac{2^{3s+1}}{s!}a(\vec{p}_1,\vec{p}_2)\label{physical primary operator2}
\end{eqnarray}
and $\mathcal{H}_{s}^{(-)}(t;\vec{x},z)$ is obtained from $\mathcal{H}_{s}^{(+)}(t;\vec{x},z)$ by replacing $J_{-\frac{1}{2}} (\sqrt{(p^0)^2-\vec{p}^2}\; z) [\widetilde{\mathcal{O}}_s(p;\epsilon)+\widetilde{\mathcal{O}}_s(p;\epsilon^*)]$ with $J_{\frac{1}{2}} (\sqrt{(p^0)^2-\vec{p}^2}\; z) {1\over i}[\widetilde{\mathcal{O}}_s(p;\epsilon)-\widetilde{\mathcal{O}}_s(p;\epsilon^*)]$.
These can be recognized as physical projections of conserved higher spin currents characterizing the UV theory.

The boundary behavior of $\mathcal{H}_s^{(\pm)}\left(t,\vec{x},z\right)$ is
\begin{eqnarray}
\fl\mathcal{H}_s^{(+)}(t,\vec{x},z)\stackrel{z\rightarrow 0}{\longrightarrow}&& 2^{-\frac{3}{2}s-1}s!\left(-(\partial^0)^2+\vec{\partial}^2\right)^{-\frac{s}{2}}\int_{p^0>|\vec{p}|} \frac{d^2\vec{p}dp^0}{(2\pi)^32p^0}\; \left[\widetilde{\mathcal{O}}_s(p;\epsilon)+\widetilde{\mathcal{O}}_s(p;\epsilon^*)\right]\cr
&&+\mbox{h.c.}
\end{eqnarray}
\begin{eqnarray}
\fl
\mathcal{H}_s^{(-)}(t,\vec{x},z)\stackrel{z\rightarrow 0}{\longrightarrow} &&  2^{-\frac{3}{2}s-1}s!z\left(-(\partial^0)^2+\vec{\partial}^2\right)^{-\frac{s}{2}+1}\int_{p^0>|\vec{p}|} \frac{d^2\vec{p}dp^0}{(2\pi)^32p^0}\; \left[\widetilde{\mathcal{O}}_s(p;\epsilon)-\widetilde{\mathcal{O}}_s(p;\epsilon^*)\right]\cr
&&+\mbox{h.c.}\quad
\end{eqnarray}

We note that for the special $s=0$ case, there exists only one scalar field $\mathcal{H}_{0}^{(+)}(t;\vec{x},z)$ because $\mathcal{H}_0^{(-)}(t;\vec{x},z)$ vanishes. On the other hand, there are two physical higher spin fields for each even spin $s$. Note that the symmetry of the bi-local field $\widetilde{\Psi}'(\vec{p}_1,\vec{p}_2)=\widetilde{\Psi}'(\vec{p}_2,\vec{p}_1)$ leads to a symmetry of the higher spin field $\widetilde{\mathcal{H}}'(\vec{p},p^z,\theta)=\widetilde{\mathcal{H}}'(\vec{p},-p^z,-\theta)$ via kernel $\mathcal{K}$. Therefore, one has two modes, $\cos p^z z\cos s\theta$ and $\sin p^z z\sin s\theta$, which corresponds to $\mathcal{H}^{(+)}_s$ and $\mathcal{H}^{(-)}_s$, respectively.

\subsection{Reduction}
\label{sec:reduction}

To confirm agreement with tensor structure of higher spin currents (and similarly with static gauge fixing in AdS) we have to perform a reduction to the physical space which we now present. This reduction follows an analogous reduction performed by Metsaev in the light-cone case to reach the independent higher spin observables. Agreement of such a reduction to physical subspace, with the collective construction, represents a most relevant verification of the collective construction for higher spins.

One packages a spin-$s$ conformal operator with oscillators~$\alpha$'s.
\begin{equation}
\mathcal{O}_s(x;\alpha)=\mathcal{O}^s_{\mu_1\mu_2\cdots \mu_s}\alpha^{\mu_1}\cdots \alpha^{\mu_s}
\end{equation}
$\mathcal{O}_s(x;\alpha)$ satisfies two equations, the current conservation and traceless condition.
\begin{eqnarray}
\bar{\alpha}^\mu\partial_\mu \mathcal{O}_s(x;\alpha)&=&0\label{conservation2nd}\\
\bar{\alpha}^\mu\bar{\alpha}_\mu  \mathcal{O}_s(x;\alpha)&=&0\label{traceless2nd}\\
\alpha^\mu\bar{\alpha}_\mu  \mathcal{O}_s(x;\alpha)&=&s\mathcal{O}^s\left(x;\alpha\right)\label{spin2nd}
\end{eqnarray}
 The last one corresponds to spin $s$ condition. To simplify constraints, we introduce $\mathcal{M}_{tot}$.
\begin{equation}
\mathcal{M}_{tot}\equiv \mathcal{M}_1\mathcal{M}_2\mathcal{M}_3\mathcal{M}_4 
\end{equation}
where
\begin{eqnarray}
\mathcal{M}_1&\equiv&\exp\left[-\alpha^0\left(\frac{1}{p^0}\left(\bar{\alpha}^1p^1+\bar{\alpha}^2p^2\right)\right)\right]\\
\mathcal{M}_2&\equiv&\exp\left[-\theta\left(\alpha^1\bar{\alpha}^2-\alpha^2\bar{\alpha}^1\right)\right]\qquad(\mbox{where}\;\; \tan\theta=p_2/ p_1)\\
\mathcal{M}_3&\equiv&\exp\left[-\alpha^1\bar{\alpha}^1\log \left(\frac{q}{ p^0}\right)\right]\\
\mathcal{M}_4&\equiv&\exp\left[\frac{\pi}{2}i\alpha^2\bar{\alpha}^2\right]\exp\left[\frac{\pi}{4}\left(\alpha^1\bar{\alpha}^2-\alpha^2\bar{\alpha}^1\right)\right]
\end{eqnarray}
$\mathcal{M}_{tot}$ is motivated by canonical transformation in the \ref{sec:polarization}. Using $\mathcal{M}_{tot}$, we rescale $\widetilde{\mathcal{O}}_s(p;\alpha)$.
\begin{equation}
\widetilde{\mathcal{O}}_s(p;\alpha)=\mathcal{M}_{tot}\mathcal{B}_s(p;\alpha)
\end{equation}
where $\widetilde{\mathcal{O}}_s(p;\alpha)$ is Fourier transformation of $\mathcal{O}_s(x;\alpha)$ to momentum space. Then, (\ref{conservation2nd})$\sim$(\ref{spin2nd}) become three constraints for $\mathcal{B}^s(p;\alpha)$.
\begin{eqnarray}
(-\bar{\alpha}^0p^0)\mathcal{B}_s(p;\alpha)=0\\
 (2\bar{\alpha}^1\bar{\alpha}^2-\bar{\alpha}^0f(p;\bar{\alpha}))\mathcal{B}_s(p;\alpha)=0\\
\alpha^\mu\bar{\alpha}_\mu  \mathcal{B}_s(p;\alpha)=\; s\mathcal{B}_s(p;\alpha)
\end{eqnarray}
Here, $f(p;\bar{\alpha})$ is a fixed function followed from the constraint (\ref{traceless2nd}). Its explicit form will not be needed below. These three equations have two independent solutions of spin $s$.
\begin{eqnarray}
\mathcal{B}_s^{(1)}(p;\alpha^1)&=&\mathcal{B}_s^{(1)}(p)\;(\alpha^1)^s\\
\mathcal{B}_s^{(2)}(p;\alpha^2)&=&\mathcal{B}_s^{(2)}(p)\;(\alpha^2)^s
\end{eqnarray}
Hence, using $\mathcal{M}_{tot}$, one can find a solution of (\ref{conservation2nd})$\sim$(\ref{spin2nd}).
\begin{eqnarray}
\widetilde{\mathcal{O}}_s(p;\alpha)&=&\mathcal{M}_{tot}\left(\mathcal{B}_s^{(1)}(p)\;(\alpha^1)^s+\mathcal{B}_s^{(2)}(p)\;(\alpha^2)^s\right)\cr
&=&\mathcal{B}_s^{(1)}(p)\left(\epsilon_\mu \alpha^\mu\right)^s+\mathcal{B}_s^{(2)}(p)\left(\epsilon^*_\mu \alpha^\mu\right)^s
\end{eqnarray}
where $\epsilon(p)$ is a polarization vector. 
\begin{equation}
\epsilon(p)\!\equiv\!\frac{1}{\sqrt{2}\left|\vec{p}\right|}\!\left(\frac{\vec{p}^2}{\sqrt{-p^\mu p_\mu}},\frac{p^0p^1}{\sqrt{-p^\mu p_\mu}}+ip^2,\frac{p^0p^2}{\sqrt{-p^\mu p_\mu}}-i p^1\right)\label{polarization2}
\end{equation}
Note that the polarization vectors $\epsilon$ and $\epsilon^*$ are null and orthogonal to $p$.
\begin{eqnarray}
\epsilon^\mu(p) \epsilon_\mu(p)&=&\epsilon^{*\mu}(p) \epsilon^*_\mu(p)=0\label{null condition}\\
p^\mu\epsilon_\mu(p)&=&p^\mu\epsilon^*_\mu(p)=0\label{orthogonal condition}\\
\epsilon^\mu(p) \epsilon^*_\mu(p)&=&1\label{normalization}
\end{eqnarray}
To express the physical operators $\mathcal{B}_s^{(a)}$ $(a=1,2)$ in terms of $\widetilde{\mathcal{O}}_s$ we identify $\alpha$ with $\epsilon(p)$ and $\epsilon^*(p)$, respectively. Then, by (\ref{null condition}) and (\ref{normalization}), one obtains
\begin{equation}
\mathcal{B}_s^{(1)}(p)=\widetilde{\mathcal{O}}_s(p;\epsilon^*)\;,\quad\mathcal{B}_s^{(2)}(p)=\widetilde{\mathcal{O}}_s(p;\epsilon)\label{physical operator form}
\end{equation}
Hence, the general solution can be expressed as
\begin{equation}
\widetilde{\mathcal{O}}_s(p;\alpha)=\widetilde{\mathcal{O}}_s(p;\epsilon^*)\left(\epsilon_\mu \alpha^\mu\right)^s+\widetilde{\mathcal{O}}_s(p;\epsilon)\left(\epsilon^*_\mu \alpha^\mu\right)^s
\end{equation}
Or, one can obtain this solution in a simpler way. $\widetilde{\mathcal{O}}_s(p;\alpha)$ can be written as
\begin{equation}
\widetilde{\mathcal{O}}_s(p;\alpha)=\widetilde{\mathcal{O}}^s_{\mu_1\mu_2\cdots\mu_s}\left(p\right)\alpha^{\mu_1}\cdots \alpha^{\mu_s}\label{eq for simpler way}
\end{equation}
Recalling (\ref{null condition})$\sim$(\ref{normalization}), one can write flat metric in terms of $\epsilon$ and $\epsilon^*$ by completeness relation.
\begin{equation}
\eta^{\mu\nu}={p_\mu p_\nu\over p^\lambda p_\lambda}+\epsilon_\mu\epsilon^*_\nu+\epsilon^*_\mu\epsilon_\nu
\end{equation}
Then, we insert the completeness relation in each contraction between $\widetilde{\mathcal{O}}^s_{\mu_1\mu_2\cdots\mu_s}$ and $\alpha$'s in (\ref{eq for simpler way}). Because $\widetilde{\mathcal{O}}^s_{\mu_1\mu_2\cdots\mu_s}(p)$ and $\epsilon$ is orthogonal to $p$, all terms related to $\hat{p}$ vanish. In addition, because of (\ref{conservation2nd}) and (\ref{traceless2nd}), $\epsilon_\mu\alpha^\mu$ cannot be mixed with $\epsilon_\mu^* \alpha^\mu$ in each term. Therefore, we have  
\begin{eqnarray}
\!\!\widetilde{\mathcal{O}}_s(p;\alpha)&=&\widetilde{\mathcal{O}}_s(p;\epsilon^*)\left(\epsilon_\mu\alpha^\mu\right)^s+\widetilde{\mathcal{O}}_s(p;\epsilon)\left(\epsilon_\mu^* \alpha^\mu\right)^s\label{conserved current}
\end{eqnarray}
which is exactly same as before. 

Now, we will obtain the detailed form of the physical primary operators $\widetilde{\mathcal{O}}_s(p;\epsilon^*)$ and $\widetilde{\mathcal{O}}_s(p;\epsilon^*)$. Useful expressions for primary currents of the free $O(N)$ vector model were given in \cite{Makeenko:1980bh,Mikhailov:2002bp,Braun:2003rp,Giombi:2009wh,Hu:2013hha}. Or, using $SO(2,3)$ algebra only, one can obtain Clebsch-Gordan coefficients for $SO(2,3)$, and can derive the same result.
\begin{eqnarray}
\hspace*{-2cm}
\mathcal{O}_s(x;\alpha)=\sum_{n=0}^{s/2}\frac{(-4)^n}{(2n)!}\sum_{k=0}^{s-2n}\frac{(-1)^k}{k!(s-2n-k)!}:(\alpha\cdot\partial)^{n+k}\phi^i(x)\;(\alpha\cdot\partial)^{s-n-k}\phi^i(x) :
\end{eqnarray}
Fourier transformation gives
\begin{eqnarray}
&&\widetilde{\mathcal{O}}_{s}(p;\alpha)=2p^0\int d^3x\; e^{-ip\cdot x}\mathcal{O}^s(x;\alpha)\cr
&=&\int d\vec{p}_1 d\vec{p}_2\; 2\mathcal{J}(\vec{p}_1,\vec{p}_2)\delta^{(3)}\left(p_1^\mu+p_2^\mu-p^\mu\right) \frac{s!}{\Gamma(s+\frac{1}{2})}\cr
&&\times (\alpha\cdot p_1+\alpha\cdot p_2)^sP_s^{-\frac{1}{2},-\frac{1}{2}}\left(\frac{\alpha\cdot p_1-\alpha\cdot p_2}{\alpha\cdot p_1+\alpha\cdot p_2}\right)a(\vec{p}_1,\vec{p}_2)\qquad
\end{eqnarray}
where $p^0>0$ and $p_{a}^\mu=(\left|\vec{p}_a\right|,\vec{p}_a)$ for $a=1,2$. For physical operator (see (\ref{physical operator form})), by inserting $\epsilon(\vec{p})$ and $\epsilon^*(\vec{p})$ into $\alpha$, one can obtain the same formula as (\ref{physical primary operator1}) and (\ref{physical primary operator2}), respectively. 

After manipulating, one can also express the physical operators as
\begin{eqnarray}
\widetilde{\mathcal{O}}_s(p;\epsilon)&=&\int d\vec{p}_1 d\vec{p}_2\; \delta^{(3)}\left(p_1^\mu+p_2^\mu-p^\mu\right)\frac{(-1)^s 2^{\frac{3}{2}s-1}}{s!}  \cr
&&\times \left(2\left|\vec{p}_1\right|\left|\vec{p}_2\right|-2\vec{p}_1\cdot \vec{p}_2\right)^{\frac{s}{2}} e^{is\Theta(\vec{p}_1,\vec{p}_2)} a(p_1,p_2)\label{physical operator1}\\
\widetilde{\mathcal{O}}_s(p;\epsilon^*)&=&\int d\vec{p}_1 d\vec{p}_2\; \delta^{(3)}(p_1^\mu+p_2^\mu-p^\mu) \frac{(-1)^s 2^{\frac{3}{2}s-1}}{s!}\cr
&&\times \left(2\left|\vec{p}_1\right|\left|\vec{p}_2\right|-2\vec{p}_1\cdot \vec{p}_2\right)^{\frac{s}{2}} e^{-is\Theta(\vec{p}_1,\vec{p}_2)}a(p_1,p_2)\label{physical operator2} 
\end{eqnarray}
where $\Theta(\vec{p}_1,\vec{p}_2)$ is the same (up to sign) as the ansatz (\ref{ansatz for theta}) for the identification of $\theta$ coordinate with bi-local momenta in the \ref{sec:time-like frame}. In fact, one can naturally derive the identification of $\theta$ from the primary operator in this way.\footnote{In the light-cone case, one can also derive the identification of $\theta$ in the similar way.}

\section{Conclusions}
\label{sec:conclusions}

In this paper, we have summarized (and elaborated further on) the canonical properties of the collective field construction of Vector Model/Higher Spin Duality. In particular, in time-like quantization a Map between bi-local collective fields and Higher Spin fields is established first. As we have explained here, one has the gauge 'reduced' canonical Higher Spin fields which for all spins are summarized as a scalar field on AdS$_4 \times S^1$, a space on which a massless representation of $SO(2,3)$ can be specified. It is then demonstrated that the Map implies canonical commutation relations for the constructed AdS fields. These relations are exact to all orders in $1/N$ and are seen to follow from the canonical property of the collective fields. As such the relations do not need modification or improvement in each order of the Large $N$ expansion. This is one of the central properties of the collective construction. As such the present exact construction might be of relevance in the study of horizon properties of Black Hole states following recent consideration of \cite{Hamilton:2006fh,Papadodimas:2012aq,Heemskerk:2012mn}. It will also be interesting to apple the canonical-bulk construction to Dualties involving Minimal Models\cite{Jevicki:2014mfa}.Finally we would like to stress that the collective constructions are exact in the perturbative sense defined by the $1/N$ expansion. In construction from CFT, one also encounters another issue which comes under the name of the `finite $N$ exclusion' \cite{Maldacena:1998bw}. In the collective approach, this represents the fact that the collective field variables represent an overcomplete set at finite $N$. Consequences of this on the emergent AdS space time were already considered in \cite{Jevicki:1998rr} and recently in \cite{Kabat:2014kfa}. In the present Vector model case an elegant Hilbert space implementation of the `exclusion principle' is presented in \cite{Das:2012dt}.

\ack
The work of AJ and JY is supported by the Department of Energy under contract DE-FG-02-91ER40688. AJ also acknowledges the hospitality of Center for Theoretical Physics, University of Wittwatersrand where this work was done. The work of JP Rodrigues is based on research supported in part by the National Research Foundation of South Africa (Grant specific unique reference number (UID 85974)). RdMK is supported by the South African Research Chairs Initiative of the Department of Science and Technology and National Research Foundation.

\appendix

\section{Algebra of Bi-local Operators}\label{sec:algebra of bi-local operators}

The bi-local operators in the light-cone case obey an algebra which can be used to specify completely their $1/N$ expansion. We explain it in the case of the $U(N)$ vector model for simplicity. Note that $O(N)$ case was already discussed in \cite{deMelloKoch:2012vc}. For $U(N)$, Fourier transformation gives two oscillators, $a_i(p^+,p)$ and $b_i(p^+,p)$ satisfying
\begin{equation}
\hspace*{-1cm}
[a_i(p^+_1,p_1),a_j^\dag (p^+_2,p_2)]=[b_i(p^+_1,p_1),b_j^\dag (p^+_2,p_2)]=\delta_{i,j}\delta(p^+_1-p^+_2)\delta(p_1-p_2)
\end{equation}
where $p^+,q^+>0$. From now on, we will suppress $p$ dependence and will write $a_i(p^+_I,p_I)$ as $a_i(I)$ for simplicity. One can construct bi-local operators including creation operators for singlet sector.
\begin{eqnarray}
A(I,J)={1\over \sqrt{N}} \sum_{i=1}^N b_i(I) a_i(J)\quad,\quad A^\dag(I,J)={1\over \sqrt{N}} \sum_{i=1}^N  a_i^\dag(I)b_i^\dag(J)\cr
B(I,J)= \sum_{i=1}^N a_i^\dag(I) a_i(J)\quad,\quad C(I,J)=\sum_{i=1}^N b_i^\dag(J) b_i(I)
\end{eqnarray}
These bi-local operators close a algebra. i.e.
\begin{eqnarray}
[B(I,J),A(K,L)]=-\delta_{I,L} A(K,J)\quad,\quad   [B(I,J),A^\dag(K,L)]=\delta_{J,K} A^\dag(I,L)\cr
[C(I,J),A(K,L)]=-\delta_{J,K} A(I,L)  \quad,\quad [C(I,J),A^\dag(K,L)]=\delta_{I,L} A^\dag(K,J)\cr
[A(I,J),A^\dag(K,L)]=\delta_{J,K}\delta_{1,L}+{1\over N} (\delta_{I,L} B(K,J)+\delta_{J,K} C(I,L))
\end{eqnarray}
In singlet sector, one can find Casimir constraints 
\begin{equation}
A^\dag(I,J)\star A(J,K)-B(I,K)-{1\over N} B(I,J)\star B(J,K) +{\delta(0)\over N} B(I,K)=0
\end{equation}
This constraint implies that the bi-local operators are not independent. To solve the constraint, we find three equivalent realizations of this algebra in terms of an bi-local oscillator $\alpha(I,J)$ satisfying $[\alpha(I,J),\alpha^\dag(K,L)]=\delta_{I,L} \delta_{J,K}$.
\begin{eqnarray}
&({\romannumeral 1})&\; \left\{
\begin{array}{l}
A(I,J)=\alpha(I,J)+{1\over N} \alpha^\dag(K,L)\alpha(I,K)\alpha(L,J)  \\
A^\dag(I,J)=\alpha^\dag(I,J)  \\
B(I,J)=\alpha^\dag(I,K)\alpha(K,J) \\
C(I,J)=\alpha^\dag(K,J)\alpha(I,K)
\end{array}\right.\\
&({\romannumeral 2})&\; \left\{
\begin{array}{l}
A(I,J)=\sqrt{\mathbb{I}+{1\over N} (\alpha\star \alpha^\dag -\delta(0) \mathbb{I})}(I,K)\star \alpha(K,J)\\
A^\dag(I,J)=\alpha^\dag(I,K)\star \sqrt{\mathbb{I}+{1\over N} (\alpha\star \alpha^\dag -\delta(0) \mathbb{I})}(K,J)\\
B(I,J)=\alpha^\dag(I,K)\alpha(K,J)\\
C(I,J)=\alpha^\dag(K,J)\alpha(I,K)
\end{array}\right.\\
&({\romannumeral 3})&\; \left\{
\begin{array}{l}
A(I,J)=\alpha(I,J)\\
A^\dag(I,J)=\alpha^\dag(I,J)+{1\over N} \alpha^\dag(I,K)\alpha^\dag(L,J)\alpha(K,L)\\
B(I,J)=\alpha^\dag(I,K)\alpha(K,J)\\
C(I,J)=\alpha^\dag(K,J)\alpha(I,K)
\end{array}\right.
\end{eqnarray}
Note that we used $O(N)$ version of the realization $({\romannumeral 1})$ in (\ref{eq:psidag}). In \cite{deMelloKoch:2012vc}, a realization similar to $({\romannumeral 2})$ was found for the case of $O(N)$. Moreover, the realization $({\romannumeral 3})$ agrees with \cite{Mintun:2014gua}.

\section{The Inverse Transformation of Momentum Space in Time-like gauge}\label{sec:inverse transformation}

\begin{eqnarray}
\fl
\beta_1^1(\vec{p},p^z,\theta)&=&\frac{\omega-|\vec{p}|\cos\theta}{2|\vec{p}|\left((p^z)^2+\vec{p}^2\sin^2\theta\right)}\cr
\fl&&\times\left[\left(\omega\left|\vec{p}\right|\sin^2\theta-(p^z)^2\cos\theta\right)p^1-\left(p^z|\vec{p}|\sin\theta\cos\theta+p^z\omega\sin\theta\right)p^2\right]\label{time-like transf1}\\
\fl\beta_1^2(\vec{p},p^z,\theta)&=&\frac{\omega-|\vec{p}|\cos\theta}{2|\vec{p}|\left((p^z)^2+\vec{p}^2\sin^2\theta\right)}\cr
\fl&&\times\left[\left(p^z|\vec{p}|\sin\theta\cos\theta+p^z\omega\sin\theta\right)p^1+\left(\omega |\vec{p} |\sin^2\theta-(p^z)^2\cos\theta\right)p^2\right]\label{time-like transf2}\\
\fl\beta_2^1(\vec{p},p^z,\theta)&=&\frac{\omega+|\vec{p}|\cos\theta}{2|\vec{p}|\left((p^z)^2+\vec{p}^2\sin^2\theta\right)}\cr
\fl&&\times\left[\left(\omega |\vec{p} |\sin^2\theta+(p^z)^2\cos\theta\right)p^1-\left(p^z |\vec{p} |\sin\theta\cos\theta-p^z\omega\sin\theta\right)p^2\right]
\label{time-like transf3}\\
\fl\beta_2^2(\vec{p},p^z,\theta)&=&\frac{\omega+|\vec{p}|\cos\theta}{2|\vec{p} |\left( (p^z )^2+\vec{p}^2\sin^2\theta\right)}\cr
\fl&&\times\left[\left(p^z |\vec{p}|\sin\theta\cos\theta-p^z\omega\sin\theta\right)p^1 +\left(\omega |\vec{p} |\sin^2\theta+ (p^z)^2\cos\theta\right)p^2\right]
\label{time-like transf4}
\end{eqnarray}
where $\omega\equiv\sqrt{(p^z)^2+\vec{p}^2}$.

\section{Polarization vector $\epsilon$}\label{sec:polarization}

Consider $\left(6+6\right)$ dimensional phase space.
\begin{equation}
(x^\mu,p^\mu;\alpha^\mu,\bar{\alpha}^\mu)\qquad \mu=0,1,2
\end{equation}
with
\begin{equation}
\left[ p^\mu,x^\nu\right]=\eta^{\mu\nu}\quad,\qquad\left[\bar{\alpha}^\mu,\alpha^\nu\right]=\eta^{\mu\nu}
\end{equation}
and others vanish. We have two constraints, conservation law and traceless condition.
\begin{eqnarray}
p^\mu\bar {\alpha}_\mu&=&0\\
\bar{\alpha}^\mu \bar{\alpha}_\mu&=&0
\end{eqnarray}
To simplify two constraints, we introduce two canonical transformation $\mathcal{T}_1$ and $\mathcal{T}_2$.
\begin{equation}
(x^\mu,p^\mu;\alpha^\mu,\bar{\alpha}^\mu)  \stackrel{\mathcal{T}_1}{\longrightarrow}  (\bar{x}^\mu,\bar{p}^\mu;\beta^\mu,\bar{\beta}^\mu) \stackrel{\mathcal{T}_2}{\longrightarrow}  (\widetilde{x}^\mu,\widetilde{p}^\mu;\gamma^\mu,\bar{\gamma}^\mu)
\end{equation}
where the canonical transformation $\mathcal{T}_1$ is defined to be
\begin{eqnarray}
x^0&=&\bar{x}^0+\frac{\beta^0}{\left(\bar{p}^0\right)^2}\left(\bar{\beta}^1\bar{p}^1+\bar{\beta}^2\bar{p}^2\right)\\
x^i&=&\bar{x}^i+\frac{1}{\bar{p}^0}\beta^0\bar{\beta}^i\\
p^\mu&=&\bar{p}^\mu\\
\alpha^0&=&\beta^0\\
\alpha^i&=&\beta^i+\frac{\bar{p}^i}{\bar{p}^0}\beta^0\\
\bar{\alpha}^0&=&\bar{\beta}^0+\frac{1}{p^0}\left(\bar{\beta}^1\bar{p}^1+\bar{\beta}^2\bar{p}^2\right)\\
\bar{\alpha}^i&=&\bar{\beta}^i
\end{eqnarray}
and the canonical transformation $\mathcal{T}_2$ is given by
\begin{eqnarray}
\bar{x}^0&=&\widetilde{x}^0-\left(\frac{\widetilde{p}^0}{2\widetilde{q}^2}+\frac{1}{2\widetilde{p}^0}\right)(\gamma^1+\gamma^2)(\bar{\gamma}^1+\bar{\gamma}^2)\\
\bar{x}^1&=&\widetilde{x}^1-\frac{\widetilde{p}^1}{2\widetilde{q}^2}(\gamma^1+\gamma^2)(\bar{\gamma}^1+\bar{\gamma}^2)+\frac{i\widetilde{p}^2}{2\widetilde{k}^2}F\\
\bar{x}^2&=&\widetilde{x}^2-\frac{\widetilde{p}^2}{2\widetilde{q}^2}(\gamma^1+\gamma^2)(\bar{\gamma}^1+\bar{\gamma}^2)-\frac{i\widetilde{p}^1}{2\widetilde{k}^2}F\\
\bar{p}^\mu&=&\widetilde{p}^\mu\\
\beta^0&=&\gamma^0\\
\beta^1&=&\frac{1}{\sqrt{2}\widetilde{k}}\left[\left(\frac{\widetilde{q}}{\widetilde{p}^0}\widetilde{p}^1-i\widetilde{p}^2\right)\gamma^1+\left(\frac{\widetilde{q}}{\widetilde{p}^0}\widetilde{p}^1+i\widetilde{p}^2 \right)\gamma^2\right]\qquad\\
\beta^2&=&\frac{1}{\sqrt{2}\widetilde{k}}\left[\left(\frac{\widetilde{q}}{\widetilde{p}^0}\widetilde{p}^2+i\widetilde{p}^1\right)\gamma^1+\left(\frac{\widetilde{q}}{\widetilde{p}^0}\widetilde{p}^2-i\widetilde{p}^1 \right)\gamma^2\right]\\
\bar{\beta}^0&=&\bar{\gamma}^0\\
\bar{\beta}^1&=&\frac{1}{\sqrt{2}\widetilde{k}}\left[\left(\frac{\widetilde{p}^0\widetilde{p}^1}{\widetilde{q}}+i\widetilde{p}^2\right)\bar{\gamma}^1+\left(\frac{\widetilde{p}^0\widetilde{p}^1}{\widetilde{q}}-i\widetilde{p}^2\right)\bar{\gamma}^2\right]\\
\bar{\beta}^2&=&\frac{1}{\sqrt{2}\widetilde{k}}\left[\left(\frac{\widetilde{p}^0\widetilde{p}^2}{\widetilde{q}}-i\widetilde{p}^1\right)\bar{\gamma}^1+\left(\frac{\widetilde{p}^0\widetilde{p}^2}{\widetilde{q}}+i\widetilde{p}^1 \right)\bar{\gamma}^2\right]
\end{eqnarray}
where $\widetilde{q}\equiv\sqrt{-\widetilde{p}^\mu\widetilde{p}_\mu}$, $\widetilde{k}\equiv \sqrt{\widetilde{p}^i\widetilde{p}^i}$ and
\begin{equation}
F(\widetilde{x}^\mu,\widetilde{p}^\mu,\gamma^\mu,\bar{\gamma}^\mu)\equiv\frac{\widetilde{p}^0(\gamma^1-\gamma^2)(\bar{\gamma}^1+\bar{\gamma}^2)}{\widetilde{q}}+\frac{\widetilde{q}(\gamma^1+\gamma^2)(\bar{\gamma}^1-\bar{\gamma}^2)}{\widetilde{p}^0}\qquad
\end{equation}
Under $\mathcal{T}_1$ and $\mathcal{T}_2$, the conservation law is transformed as follows.
\begin{equation}
\bar{\alpha}^\mu p_\mu=0 \stackrel{\mathcal{T}_1}{\longrightarrow}  -\bar{\beta}^0\bar{p}^0=0  \stackrel{\mathcal{T}_2}{\longrightarrow}  -\bar{\gamma}^0\widetilde{p}^0=0\label{mod constraints1}
\end{equation}
Moreover, the transformation of the traceless condition is
\begin{eqnarray}
\bar{\alpha}^\mu\bar{\alpha}_\mu=0\;&&\stackrel{\mathcal{T}_1}{\longrightarrow} \;(\bar{\beta}^1)^2+(\bar{\beta}^2)^2-\frac{1}{\left(\bar{p}^0\right)^2}(\bar{\beta}^0\bar{p}^0+\bar{\beta}^1\bar{p}^1+\bar{\beta}^2\bar{p}^2)^2=0\cr
&& \stackrel{\mathcal{T}_2}{\longrightarrow} \; 2\bar{\gamma}^1\bar{\gamma}^2-\bar{\gamma}^0f(\widetilde{p},\bar{\gamma})=0\label{mod constraints2}
\end{eqnarray}
where $f\left(\tilde{p},\bar{\gamma}\right)$ is an unimportant function of $\widetilde{p}^\mu$ and $\bar{\gamma}$.

One can easily solve (\ref{mod constraints1}) and (\ref{mod constraints2}). The first solution is
\begin{equation}
\gamma^0=\bar{\gamma}^0=\gamma^2=\bar{\gamma}^2=0
\end{equation}
Then, the remaining dynamical variables are $\left(\widetilde{x};\gamma^1\right)$. Especially, one can inverse transform $\gamma^1$ into original phase space.
\begin{equation}
\gamma^1\quad  \stackrel{{\mathcal{T}^{-1}_1\circ\mathcal{T}^{-1}_2}}{\longrightarrow} \quad\epsilon_\mu(p) \alpha^\mu
\end{equation}
where $\epsilon(p)$ is the null polarization vector found in (\ref{polarization2}).
\begin{equation}
\epsilon(p)\!=\!\frac{1}{\sqrt{2}\left|\vec{p}\right|}\!\left(\frac{\vec{p}^2}{\sqrt{-p^\mu p_\mu}},\frac{p^0p^1}{\sqrt{-p^\mu p_\mu}}+ip^2,\frac{p^0p^2}{\sqrt{-p^\mu p_\mu}}-i p^1\right)\label{polarization}
\end{equation}
There is another solution of (\ref{mod constraints1}) and (\ref{mod constraints2}).
\begin{equation}
\gamma^0=\bar{\gamma}^0=\gamma^1=\bar{\gamma}^1=0
\end{equation}
Dynamical variables are $\left(\widetilde{x};\gamma^2\right)$ and the inverse transformation of $\gamma^2$ is
\begin{equation}
\gamma^2\quad \stackrel{{\mathcal{T}^{-1}_1\circ\mathcal{T}^{-1}_2}}{\longrightarrow}  \quad\epsilon^*_\mu(p) \alpha^\mu
\end{equation}
where $\epsilon^*(p)$ is a null polarization of this solution and is complex conjugate of (\ref{polarization}).

\section{Representation of $SO\left(2,3\right)$}\label{SO(2,3) realization}

\subsection{Covariant Realization}
\label{covariant realization}

A covariant realization of $SO(2,3)$ generators acting on totally symmetric conformal operators of spin $s$ is given by \cite{Metsaev:1999ui}.
\begin{eqnarray}
P^a&=&\partial^a\\
J^{ab}&=&x^a\partial^b-x^b\partial^a+M^{ab}\\
D&=&x^a\partial_a+\Delta\\
K^a&=&-\frac{1}{2}x^b x_b \partial^a+x^aD+M^{ab}x_b
\end{eqnarray}
with
\begin{equation}
M^{ab}=\alpha^a\bar{\alpha}^b-\alpha^b\bar{\alpha}^a\quad,\qquad \Delta=s+1
\end{equation}
The condition $\Delta=s+1$ is required because the conservation equation needs to commute with all $SO(2,3)$ generators (especially, $K^\mu$).

\subsection{Realization on Physical Operator $\widetilde{\mathcal{O}}(p^0,\vec{p})$}
\label{sec:Realization on Physical Operator1}

In \ref{sec:projection to currents}, we obtained the physical operator by solving the conservation law and traceless condition. Hence, using $\mathcal{M}_{tot}$ again, one can find realization of $SO\left(2,3\right)$ acting on the physical operator $\widetilde{\mathcal{O}}_s(p;\epsilon^*)$. i.e.
\begin{equation}
L^{AB}\widetilde{\mathcal{O}}_s(p;\alpha)=\mathcal{M}_{tot}\bar{L}^{AB}\widetilde{\mathcal{O}}_s(p;\epsilon^*)
\end{equation}
where $L^{AB}$ is the covariant realization of $SO\left(2,3\right)$ in the \ref{covariant realization} and $\bar{L}^{AB}$ acts on the physical operator $\widetilde{\mathcal{O}}_s (p;\epsilon^* )$. The result is
\begin{eqnarray}
\hspace*{-2cm}
\bar{P}^0&=&p^0\\
\hspace*{-2cm}
\bar{P}^1&=&p^1\\
\hspace*{-2cm}
\bar{P}^2&=&p^2\\
\hspace*{-2cm}
\bar{D}&=&-x^0p^0+x^1p^1+x^2p^2+2\alpha^1\bar{\alpha}^1\\
\hspace*{-2cm}
\bar{J}^{01}&=&x^0p^1-x^1p^0+\left(-\frac{p^1}{p^0}+i\frac{p^2}{q}-i\frac{(p^0)^2p^2}{q\vec{p}^2}\right)\alpha^1\bar{\alpha}^1\\
\hspace*{-2cm}
\bar{J}^{12}&=&x^1p^2-x^2p^1\\
\hspace*{-2cm}
\bar{J}^{20}&=&x^2p^0-x^0p^2+\left(\frac{p^2}{p^0}+i\frac{p^1}{q}-i\frac{(p^0)^2p^1}{q\vec{p}^2}\right)\alpha^1\bar{\alpha}^1\\
\hspace*{-2cm}
\bar{K}^0&=&-\frac{1}{2}(x^\mu x_\mu)p^0+x^0 \bar{D}+\frac{1}{2}(\alpha^1\bar{\alpha}^1)^2\left(\frac{q^2}{p^0\vec{p}^2}-\frac{3}{p^0}-\frac{\vec{p}^2}{p^0 q^2}\right)\cr
\hspace*{-2cm}&&-(\alpha^1\bar{\alpha}^1)\left(\frac{x^1p^1+x^2p^2}{p^0}+i\frac{q}{\vec{p}^2}(x^1p^2-x^2p^1)\right)\\
\hspace*{-2cm}
\bar{K}^1&=&-\frac{1}{2}(x^\mu x_\mu)p^1+x^1 \bar{D}+\frac{1}{2}(\alpha^1\bar{\alpha}^1)^2\left(-\frac{p^1  }{\vec{p}^2}-\frac{p^1\vec{p}^2}{q^2 p^0}-2i\frac{p^2(\vec{p}^2-2(p^0)^2)}{p^0q\vec{p}^2}\right)\cr
\hspace*{-2cm}&&-(\alpha^1\bar{\alpha}^1)\left(\frac{p^1}{p^0}+i\frac{p^2}{\vec{p}^2}q\right)x^0\\
\hspace*{-2cm}
\bar{K}^2&=&-\frac{1}{2}(x^\mu x_\mu)p^2+x^2 \bar{D}+\frac{1}{2}(\alpha^1\bar{\alpha}^1)^2\left(-\frac{p^2  }{\vec{p}^2}-\frac{p^2\vec{p}^2}{(p^0)^2 q^2}+2i\frac{p^1(\vec{p}^2-2(p^0)^2)}{p^0q\vec{p}^2}\right)\cr
\hspace*{-2cm}&&-(\alpha^1\bar{\alpha}^1)\left(\frac{p^2}{p^0}-i\frac{p^1}{\vec{p}^2}q\right)x^0
\end{eqnarray}
where $q=\sqrt{-p^\mu p_\mu}$.

\subsection{Realization on Physical Operator $\widetilde{\mathcal{O}}\left(\vec{p},p^z\right)$}
\label{sec:Realization on Physical Operator2}

One can change momentum from $(p^0,p^1,p^2)$ to $(p^1,p^2,p^z)$ by using on-shell condition
\begin{equation}
p^z=\sqrt{(p^0)^2-(p^1)^2-(p^2)^2}
\end{equation}
For coordinates, one can transform $(x^0,x^1,x^2)$ to $(x^1,x^2,z)$ by chain rule. Using $\bar{L}^{AB}$ in \ref{sec:Realization on Physical Operator1}, we obtain a representation of $SO(2,3)$ generators acting on $\widetilde{\mathcal{O}}(p^1,p^2,p^z;\epsilon^*)$
\begin{eqnarray}
P^0&=&\omega\\
P^1&=&p^1\\
P^2&=&p^2\\
D&=&x^1p^1 +x^2p^2+zp^z +2\alpha^1\bar{\alpha}^1\\
J^{01}&=&-x^1\omega-\left(\frac{p^1}{\omega}+i\frac{p^2p^z}{\vec{p}^2}\right)\alpha^1\bar{\alpha}^1\\
J^{12}&=&x^1p^2-x^2p^1\\
J^{20}&=&x^2\omega+\left(\frac{p^2}{\omega}-i\frac{p^1p^z}{\vec{p}^2}\right)\alpha^1\bar{\alpha}^1
\end{eqnarray}
\begin{eqnarray}
K^0&=&-\frac{1}{2}(\vec{x}^1+z^2)\omega-\frac{1}{2}(\alpha^1\bar{\alpha}^1)^2\left(\frac{\omega}{(p^z)^2}+\frac{3}{\omega}-\frac{\omega}{\vec{p}^2}\right)\cr
&&-(\alpha^1\bar{\alpha}^1)\left[\frac{z\omega}{p^z}+\frac{x^1p^1+x^2p^2+zp^z}{\omega}+i\frac{p^zJ^{12}}{\vec{p}^2}\right]\\
K^1&=&-\frac{1}{2}(\vec{x}^1+z^2)p^1+x^1D-\left(\frac{p^1}{p^z}-i\frac{p^2\omega}{\vec{p}^2}\right)z\alpha^1\bar{\alpha}^1\cr
&&-\frac{1}{2}(\alpha^1\bar{\alpha}^1)^2\left[\frac{p^1 \vec{p}^2}{(p^z)^2\omega^2}+\frac{p^1}{\vec{p}^2}-2i\frac{\omega p^2}{\vec{p}^2 p^z}-2i\frac{p^2p^z}{\vec{p}^2\omega}\right]\\
K^2&=&-\frac{1}{2}(\vec{x}^1+z^2)p^2+x^2D-\left(\frac{p^2}{p^z}+i\frac{p^1\omega}{\vec{p}^2}\right)z\alpha^1\bar{\alpha}^1\cr
&&-\frac{1}{2}(\alpha^1\bar{\alpha}^1)^2\left[\frac{p^2 \vec{p}^2}{(p^z)^2\omega^2}+\frac{p^2}{\vec{p}^2}+2i\frac{\omega p^1}{\vec{p}^2 p^z}+2i\frac{p^1p^z}{\vec{p}^2\omega}\right]
\end{eqnarray}
where $\omega=\sqrt{\vec{p}+(p^z)^2}$.

\subsection{Simple Realization in Different Basis}

To simplify generators, we introduce a new basis $\widetilde{\mathcal{O}}_{mod}(p;\epsilon^*(p))$
\begin{equation}
\widetilde{\mathcal{O}}_{\text{mod}} (p;\epsilon^* )=e^{-\alpha^1\bar{\alpha}^1\log\left(p^0\sqrt{(p^0)^2-\vec{p}^2}\right)}\widetilde{\mathcal{O}} (p;\epsilon^* )
\end{equation}
From the generators $\bar{L}^{AB}$ in \ref{sec:Realization on Physical Operator1}, one can find $SO\left(2,3\right)$ generators acting on $\widetilde{\mathcal{O}}_{mod} (p;\epsilon^* (p ) )$. Then, one can also change momentum from $\left(p^0,p^1,p^2\right)$ to $\left(p^1,p^2,p^z\right)$ by on-shell condition as we did in the \ref{sec:Realization on Physical Operator2}. The resulting generators $L^{ads}_{\text{mod}}$ are
\begin{eqnarray}
P^0_{\text{mod}}&=&\sqrt{\vec{p}^2+(p^z)^2}\label{mod gen1}\\
P^1_{\text{mod}}&=&p^1\label{mod gen2}\\
P^1_{\text{mod}}&=&p^2\label{mod gen3}\\
J^{01}_{\text{mod}}&=& -x^1 P^0- \frac{p^2p^z}{\vec{p}^2}(i\alpha^1 \bar{\alpha}^1 )\label{mod gen4}\\
J^{12}_{\text{mod}}&=& x^1p^2-x^2 p^1\label{mod gen5}\\
J^{20}_{\text{mod}}&=& x^2 P^0-\frac{p^1p^z}{\vec{p}^2}(i\alpha^1 \bar{\alpha}^1 )\label{mod gen6}\\
D_{\text{mod}}&=&x^1p^1+x^2p^2+z p^z\label{mod gen7}\\
K^0_{\text{mod}}&=&-\frac{1}{2}(\vec{x}^2+z^2)P^0-\frac{p^z }{\vec{p}^2}J^{12}(i\alpha^1 \bar{\alpha}^1)-\frac{P^0}{2\vec{p}^2}(i\alpha^1 \bar{\alpha}^1)^2\label{mod gen8}\\
K^1_{\text{mod}}&=&-\frac{1}{2}(\vec{x}^2+z^2)p^1+x^1D+\frac{zp^2P^0 }{\vec{p}^2}(i\alpha^1 \bar{\alpha}^1)+\frac{p^1}{2\vec{p}^2}(i\alpha^1 \bar{\alpha}^1)^2\label{mod gen9}\\
K^2_{\text{mod}}&=&-\frac{1}{2}(\vec{x}^2+z^2)p^2+x^2D-\frac{z p^1P^0 }{\vec{p}^2}(i\alpha^1 \bar{\alpha}^1)+\frac{p^2}{2\vec{p}^2}(i\alpha^1 \bar{\alpha}^1)^2\label{mod gen10}
\end{eqnarray}
where $p^\theta=i\alpha^1\bar{\alpha}^1$.

\subsection{Time-like Form of $SO\left(2,3\right)$ Generators : Candidate}

By time translation, one has a representation of $SO\left(2,3\right)$ in time-like gauge for arbitrary time $t$. i.e.
\begin{equation}
\widetilde{\mathcal{O}}(p;\epsilon^*(p))=e^{-tP^0}\widetilde{\mathcal{O}}_{\text{mod}}(p;\epsilon^*(p))
\end{equation}
Here, $t$ is not phase space variables, but a evolution parameter. The generators $L^{ab}_{\text{time-like AdS}}$ acting on $\widetilde{\mathcal{O}}\left(p;\epsilon^*\left(p\right)\right)$ are 
\begin{eqnarray}\fl
P^0&=&\sqrt{\vec{p}^2+(p^z)^2}\label{ads gen1}\\
\fl
P^1&=&p^1\label{ads gen2}\\
\fl
P^1&=&p^2\label{ads gen3}\\
\fl
J^{01}&=&t p^1-x^1 P^0- \frac{p^2p^z}{\vec{p}^2}(i\alpha^1 \bar{\alpha}^1 )\label{ads gen4}\\
\fl
J^{12}&=& x^1p^2-x^2 p^1\label{ads gen5}\\
\fl
J^{20}&=& x^2 P^0-t p^2-\frac{p^1p^z}{\vec{p}^2}(i\alpha^1 \bar{\alpha}^1 )\label{ads gen6}\\
\fl
D&=&-t\sqrt{\vec{p}^2+\left(p^z\right)^2}+x^1p^1+x^2p^2+z p^z\label{ads gen7}\\
\fl
K^0&=&-\frac{1}{2}(-t^2+\vec{x}^2+z^2)P^0+tD-\frac{p^z }{\vec{p}^2}J^{12}(i\alpha^1 \bar{\alpha}^1)-\frac{P^0}{2\vec{p}^2}(i\alpha^1 \bar{\alpha}^1)^2\label{ads gen8}\\
\fl
K^1&=&-\frac{1}{2}(-t^2+\vec{x}^2+z^2)p^1+x^1D-\frac{p^2(tp^z-zP^0) }{\vec{p}^2}(i\alpha^1 \bar{\alpha}^1)+\frac{p^1}{2\vec{p}^2}(i\alpha^1 \bar{\alpha}^1)^2\label{ads gen9}\\
\fl
K^2&=&-\frac{1}{2}(-t^2+\vec{x}^2+z^2)p^2+x^2D-\frac{p^1(zP^0-tp^z) }{\vec{p}^2}(i\alpha^1 \bar{\alpha}^1)+\frac{p^2}{2\vec{p}^2}(i\alpha^1 \bar{\alpha}^1)^2\label{ads gen10}
\end{eqnarray}
where $p^\theta=i\alpha^1\bar{\alpha}^1$. Note that one has to begin with $SO\left(2,3\right)$ generators for AdS$_4$ higher spin fields and fix time-like gauge to get precise $SO\left(2,3\right)$ generators in time-like gauge. However, we started from covariant $SO\left(2,3\right)$ realization for spin-$s$ current of CFT$_3$, and solved conservation law and traceless condition. After modification, we ended up with $L^{ab}_{\text{time-like AdS}}$. Hence, one cannot guarantee that $L^{ab}_{\text{time-like AdS}}$ is indeed a realization of $SO\left(2,3\right)$ for AdS$_4$ higher spin field in time-like gauge.

Nevertheless, there are several signs that $L^{ab}_{\text{time-like AdS}}$ is realization for AdS$_4$ higher spin field in time-like gauge. First, $L^{ab}_{\text{time-like AdS}}$ satisfy $SO\left(2,3\right)$ algebra. Second, $L^{ab}_{\text{time-like AdS}}$ for $p^\theta=0$ is equal to the realization of $SO\left(2,3\right)$ for the scalar field in AdS$_4$. Finally, when repeating the same procedure in the case of light-cone gauge, one can precisely obtain the Metsaev's light-cone gauge generators for AdS$_4$ higher spin field.

\subsection{Bi-local realization}

Bi-local realization of $SO\left(2,3\right)$ generators in time-like gauge is
\begin{eqnarray}
P^0_{(i)}&=&\sqrt{\vec{p}_{(i)}^2}\\
\vec{P}_{(i)}&=&\vec{p}_{(i)}\\
D_{(i)}&=&\vec{x}_{(i)}\cdot\vec{p}_{(i)}\\
J^{01}_{(i)}&=&-x^1_{(i)} P^0_{(i)}\\
J^{12}_{(i)}&=&x^1_{(i)} p^2_{(i)}-x^2_{(i)} p^1_{(i)}\\
J^{20}_{(i)}&=&x^2 P^0_{(i)}\\
K^0_{(i)}&=&-\frac{1}{2}(\vec{x}^2_{(i)})P^0_{(i)}\\
\vec{K}_{(i)}&=&-\frac{1}{2}(\vec{x}^2_{(i)})\vec{p}_{(i)}+\vec{x}D_{(i)}
\end{eqnarray}
where we put $t=0$ for simplicity. Then,
\begin{eqnarray}
P^\mu_{\text{bi-local}}&=&P^\mu_{(1)}+P^\mu_{(2)}\label{bi-local generator1}\\
D_{\text{bi-local}}&=&D_{(1)}+D_{(2)}\\
J_{\text{bi-local}}^{\mu\nu}&=&J^{\mu\nu}_{(1)}+J^{\mu\nu}_{(2)}\\
K_{\text{bi-local}}^\mu&=&K^\mu_{(1)}+K^\mu_{(2)}\label{bi-local generator2}
\end{eqnarray}


\section*{References}
\begin{thebibliography}{99}

\bibitem{Maldacena:1997re} 
  Maldacena~J~M 1998 The Large N limit of superconformal field theories and supergravity {\it Adv.\ Theor.\ Math.\ Phys.\ } {\bf 2} 231 (arXiv:hep-th/9711200)

\bibitem{Gubser:1998bc} 
  Gubser~S~S, Klebanov~I~R and Polyakov~A~M 1998 Gauge theory correlators from noncritical string theory {\it Phys.\ Lett.\ } B {\bf 428} 105 (arXiv:hep-th/9802109)
  
\bibitem{Witten:1998qj}
  Witten~E 1998 Anti-de Sitter space and holography {\it Adv.\ Theor.\ Math.\ Phys.\ } {\bf 2} 253 (arXiv:hep-th/9802150)

  
\bibitem{Klebanov:2002ja} 
  Klebanov~I~R and Polyakov~A~M 2002 AdS dual of the critical $O(N)$ vector model {\it Phys.\ Lett.\ }B {\bf 550} 213 (arXiv:hep-th/0210114)

\bibitem{Sezgin:2002rt} 
Sezgin~E and Sundell~P 2002 Massless higher spins and holography {\it Nucl.\ Phys.\ }B {\bf 644} 303 (arXiv:hep-th/0205131)
\item[] Sezgin~E and Sundell~P 2003 {\it Nucl.\ Phys.\ }B {\bf 660} 403 (erratum)


\bibitem{Giombi:2009wh}
  Giombi~S and Yin~X 2010 Higher Spin Gauge Theory and Holography: The Three-Point Functions {\it J.\ High Energy Phys.\ } JHEP09(2010)115 (arXiv:0912.3462~[hep-th])

\bibitem{Giombi:2013fka} 
  Giombi~S and Klebanov~I~R 2013 One Loop Tests of Higher Spin AdS/CFT {\it J.\ High Energy Phys.\ } JHEP12(2013)068 (arXiv:1308.2337~[hep-th])  

\bibitem{Jevicki:2014mfa} 
  Jevicki~A, Jin~K and Yoon~J 2014 1/N and Loop Corrections in Higher Spin AdS$_4$/CFT$_3$ Duality {\it Phys.\ Rev.\ }D {\bf 89} 085039 (arXiv:1401.3318~[hep-th])
  
\bibitem{Koch:2014mxa} 
  de~Mello~Koch~R, Jevicki~A, Rodrigues~J~P and Yoon~J 2014 Holography as a Gauge Phenomenon in Higher Spin Duality arXiv:1408.1255~[hep-th]


\bibitem{Rarita:1941mf} 
  Rarita~W and Schwinger~J 1941 On a theory of particles with half integral spin {\it Phys.\ Rev.\ }{\bf 60} 61
 
\bibitem{Fronsdal:1978rb}
  Fronsdal~C 1978 Massless Fields with Integer Spin {\it Phys.\ Rev.\ }D {\bf 18} 3624
  
\bibitem{Fang:1978wz}
  Fang~J and Fronsdal~C 1978 Massless Fields with Half Integral Spin {\it Phys.\ Rev.\ }D {\bf 18} 3630
  
\bibitem{Fradkin:1986ka} 
  Fradkin~E~S and Vasiliev~M~A 1987 Candidate to the Role of Higher Spin Symmetry {\it Annals Phys.\ }{\bf 177} 63

\bibitem{Vasiliev:1990en}
  Vasiliev~M~A 1990 Consistent equation for interacting gauge fields of all spins in (3+1)-dimensions {\it Phys.\ Lett.\ }B {\bf 243} 378

\bibitem{Vasiliev:1995dn}
  Vasiliev~M~A 1996 Higher spin gauge theories in four-dimensions, three-dimensions, and two-dimensions {\it Int.\ J.\ Mod.\ Phys.\ }D {\bf 5} 763 (arXiv:hep-th/9611024)
  
\bibitem{Bekaert:2005vh} 
  Bekaert~X, Cnockaert~S, Iazeolla~C and Vasiliev~M~A 2005 Nonlinear higher spin theories in various dimensions arXiv:hep-th/0503128

\bibitem{Das:2003vw} 
  Das~S~R and Jevicki~A 2003 Large $N$ collective fields and holography {\it Phys.\ Rev.\ }D {\bf 68} 044011 (arXiv:hep-th/0304093)

\bibitem{Koch:2010cy} 
  de Mello Koch~R, Jevicki~A, Jin~K and Rodrigues~J~P 2011 AdS$_4$/CFT$_3$ Construction from Collective Fields {\it Phys.\ Rev.\ }D {\bf 83} 025006 (arXiv:1008.0633~[hep-th])

\bibitem{Jevicki:2011ss} 
  Jevicki~A, Jin~K and Ye~Q 2011 Collective Dipole Model of AdS/CFT and Higher Spin Gravity {\it J.\ Phys.\ }A {\bf 44} 465402 (arXiv:1106.3983~[hep-th])

\bibitem{Metsaev:1999ui} 
  Metsaev~R~R 1999 Light cone form of field dynamics in Anti-de Sitter space-time and AdS / CFT correspondence {\it Nucl.\ Phys.\ }B {\bf 563} 295 (arXiv:hep-th/9906217)

\bibitem{Brodsky:2013dca} 
  Brodsky~S~J, de Teramond~G~F and Dosch~H~G 2014 QCD on the Light-Front -- A Systematic Approach to Hadron Physics {\it Few Body Syst.\ }{\bf 55} 407 (arXiv:1310.8648~[hep-ph])

\bibitem{Brodsky:2014yha}
  Brodsky~S~J, de Teramond~G~F, Dosch~H~G and Erlich~J 2014 Light-Front Holographic QCD and Emerging Confinement arXiv:1407.8131~[hep-ph]

\bibitem{Leigh:2014tza} 
  Leigh~R~G, Parrikar~O and Weiss~A~B 2014 The Holographic Geometry of the Renormalization Group and Higher Spin Symmetries {\it Phys.\ Rev.\ }D {\bf 89} 106012 (arXiv:1402.1430~[hep-th])

\bibitem{Leigh:2014qca} 
  Leigh~R~G, Parrikar~O and Weiss~A~B 2014 The Exact Renormalization Group and Higher-spin Holography arXiv:1407.4574~[hep-th]
 
\bibitem{Zayas:2013qda} 
  Pando~Zayas~L~A and Peng~C 2013 Toward a Higher-Spin Dual of Interacting Field Theories {\it J.\ High Energy Phys.\ } JHEP10(2013)023 (arXiv:1303.6641~[hep-th])


\bibitem{Douglas:2010rc} 
  Douglas~M~R, Mazzucato~L and Razamat~S~S 2011 Holographic dual of free field theory {\it Phys.\ Rev.\ }D {\bf 83} 071701 (arXiv:1011.4926~[hep-th])

\bibitem{deMelloKoch:2012vc} 
  de Mello Koch~R, Jevicki~A, Jin~K, Rodrigues~J~P and Ye~Q 2013 S=1 in O(N)/HS duality {\it Class.\ Quant.\ Grav.\ }{\bf 30} 104005 (arXiv:1205.4117~[hep-th])


\bibitem{Das:2012dt} 
  Das~D, Das~S~R, Jevicki~A and Ye~Q 2013 Bi-local Construction of Sp(2N)/dS Higher Spin Correspondence {\it J.\ High Energy Phys.\ } JHEP01(2013)107 (arXiv:1205.5776~[hep-th])

\bibitem{Metsaev:2011iz}
  Metsaev~R~R 2013 Extended Hamiltonian Action for Arbitrary Spin Fields in Flat And AdS Spaces {\it J.\ Phys.\ }A {\bf 46} 214021 (arXiv:1112.0976~[hep-th])

\bibitem{Vasiliev:2012vf} 
  Vasiliev~M~A 2013 Holography, Unfolding and Higher-Spin Theory {\it J.\ Phys.\ }A {\bf 46} 214013 (arXiv:1203.5554~[hep-th])


\bibitem{Jevicki:1979mb} 
  Jevicki~A and Sakita~B 1980 The Quantum Collective Field Method and Its Application to the Planar Limit {\it Nucl.\ Phys.\ }B {\bf 165} 511




 

  
  
\bibitem{Das:1990kaa} 
  Das~S~R and Jevicki~A 1990 String Field Theory and Physical Interpretation of $D=1$ Strings {\it Mod.\ Phys.\ Lett.\ }A {\bf 5} 1639

\bibitem{deMelloKoch:2003pv} 
  de Mello Koch~R, Donos~A, Jevicki~A and Rodrigues~J~P 2003 Derivation of string field theory from the large N BMN limit {\it Phys.\ Rev.\ }D {\bf 68} 065012 (arXiv:hep-th/0305042)








\bibitem{Bena:1999jv}
  Bena~I 2000 On the construction of local fields in the bulk of AdS(5) and other spaces {\it Phys.\ Rev.\ }D {\bf 62} 066007 (arXiv:hep-th/9905186)


\bibitem{Hamilton:2005ju}
  Hamilton~A, Kabat~D~N, Lifschytz~G and Lowe~D~A 2006 Local bulk operators in AdS/CFT: A Boundary view of horizons and locality {\it Phys.\ Rev.\ }D {\bf 73} 086003 (arXiv:hep-th/0506118)


\bibitem{Heemskerk:2012np} 
  Heemskerk~I 2012 Construction of Bulk Fields with Gauge Redundancy {\it J.\ High Energy Phys.\ } JHEP09(2012)106 (arXiv:1201.3666~[hep-th])

\bibitem{Kabat:2012hp} 
  Kabat~D, Lifschytz~G, Roy~S and Sarkar~D 2012 Holographic representation of bulk fields with spin in AdS/CFT {\it Phys.\ Rev.\ }D {\bf 86} 026004 (arXiv:1204.0126~[hep-th])
  
\bibitem{Aizawa:2014yqa} 
  Aizawa~N and Dobrev~V~K 2014 Intertwining Operator Realization of anti de Sitter Holography arXiv:1406.2129~[hep-th]
  
  
\bibitem{Fronsdal:1974ew} 
  Fronsdal~C 1974 Elementary particles in a curved space. ii {\it Phys.\ Rev.\ }D {\bf 10} 589

\bibitem{Makeenko:1980bh}
  Makeenko~Y~M 1981 Conformal Operators In Quantum Chromodynamics {\it Sov.\ J.\ Nucl.\ Phys.\ }{\bf 33} 440

\bibitem{Mikhailov:2002bp}
  Mikhailov~A 2002 Notes on higher spin symmetries arXiv:hep-th/0201019

\bibitem{Braun:2003rp}
  Braun~V~M, Korchemsky~G~P and Mueller~D 2003 The Uses of conformal symmetry in QCD {\it Prog.\ Part.\ Nucl.\ Phys.\ }{\bf 51} 311 (arXiv:hep-th/0306057)
  
\bibitem{Hu:2013hha} 
  Hu~S and Li~T 2013 Radial quantization of the 3d CFT and the higher spin/vector model duality arXiv:1312.1545~[hep-th]

\bibitem{Hamilton:2006fh}
  Hamilton~A, Kabat~D~N, Lifschytz~G and Lowe~D~A 2007 Local bulk operators in AdS/CFT: A Holographic description of the black hole interior {\it Phys.\ Rev.\ }D {\bf 75} 106001 (arXiv:hep-th/0612053)


\bibitem{Papadodimas:2012aq}
  Papadodimas~K and Raju~S 2013 An Infalling Observer in AdS/CFT {\it J.\ High Energy Phys.\ } JHEP10(2013)212 (arXiv:1211.6767~[hep-th])

\bibitem{Heemskerk:2012mn} 
  Heemskerk~I, Marolf~D, Polchinski~J and Sully~J 2012 Bulk and Transhorizon Measurements in AdS/CFT {\it J.\ High Energy Phys.\ } JHEP10(2012)165 (arXiv:1201.3664~[hep-th])

\bibitem{Maldacena:1998bw}
  Maldacena~J~M and Strominger~A 1998 AdS(3) black holes and a stringy exclusion principle {\it J.\ High Energy Phys.\ } JHEP12(1998)005 (arXiv:hep-th/9804085)

\bibitem{Jevicki:1998rr} 
  Jevicki~A and Ramgoolam~S 1999 Noncommutative gravity from the AdS / CFT correspondence {\it J.\ High Energy Phys.\ } JHEP04(1999)032 (arXiv:hep-th/9902059)

\bibitem{Kabat:2014kfa} 
  Kabat~D and Lifschytz~G 2014 Finite N and the failure of bulk locality: Black holes in AdS/CFT arXiv:1405.6394~[hep-th]



\bibitem{Mintun:2014gua} 
  Mintun~E and Polchinski~J 2014 Higher Spin Holography, RG, and the Light Cone (arXiv:1411.3151~[hep-th])


\end{thebibliography}
\end{document}